\documentclass[12pt, draftclsnofoot, onecolumn]{IEEEtran}

\usepackage{enumerate}
\usepackage{amssymb}
\usepackage{xcolor}

\usepackage{graphicx}
\usepackage{stfloats}
\ifCLASSOPTIONcompsoc
\usepackage[caption=false,font=normalsize,labelfon
t=sf,textfont=sf]{subfig}
\else
\usepackage[caption=false,font=footnotesize]{subfig}
\fi

\usepackage{amsfonts}
\usepackage{amsmath}
\usepackage{amsthm}
\usepackage{autobreak}
\usepackage{algorithm}
\usepackage{algorithmic}

\usepackage[adjust]{cite}
\usepackage[colorlinks,linkcolor=blue,citecolor=blue]{hyperref}

\usepackage{booktabs}

\usepackage{amsthm}
\theoremstyle{definition}
\newtheorem{lem}{Lemma}
\newtheorem{prop}{Proposition}

\newtheorem{rem}{Remark}

\newtheorem{deff}{Definition}

\newcounter{eqlabel}[subsection]
\newcommand{\equlabel}{\refstepcounter{eqlabel}(\alph{eqlabel})}

\begin{document}

%

\title{Impact and Calibration of Nonlinear Reciprocity Mismatch in Massive MIMO Systems}

\author{Rongjiang Nie,~Li Chen,~Nan Zhao,~\IEEEmembership{Senior Member,~IEEE},~Yunfei Chen,\\~\IEEEmembership{Senior Member,~IEEE},~Weidong Wang, and~Xianbin Wang,~\IEEEmembership{Fellow,~IEEE}
	\thanks{
		R. Nie, L. Chen and W. Wang are with the CAS Key Laboratory of Wireless Optical Communication, University of Science and Technology of China, Hefei 230052, China (e-mail: johnnrj@mail.ustc.edu.cn; \{chenli87, wdwang\}@ustc.edu.cn).}
	\thanks{N. Zhao is with the School of Info. and Commun. Eng., Dalian University of Technology, Dalian 116024, China (e-mail:zhaonan@dlut.edu.cn).}
	\thanks{Y. Chen is with the School of Engineering, University of Warwick, Coventry CV4 7AL, U.K (e-mail: Yunfei.Chen@warwick.ac.uk).}
	\thanks{X. Wang is with the Department of Electrical and Computer Engineering, Western University, London, ON N6A 3K7, Canada (e-mail: xianbin.wang@uwo.ca).}
}

\maketitle

\begin{abstract}
	Time-division-duplexing massive multiple-input multiple-output (MIMO) systems estimate the channel state information (CSI) by leveraging the uplink-downlink channel reciprocity, which is no longer valid when the mismatch arises from the asymmetric uplink and downlink radio frequency (RF) chains. Existing works treat the reciprocity mismatch as constant for simplicity. However, the practical RF chain consists of nonlinear components, which leads to nonlinear reciprocity mismatch. In this work, we examine the impact and the calibration approach of the nonlinear reciprocity mismatch in massive MIMO systems. To evaluate the impact of the nonlinear mismatch, we first derive the closed-form expression of the ergodic achievable rate. Then, we analyze the performance loss caused by the nonlinear mismatch to show that the impact of the mismatch at the base station (BS) side is much larger than that at the user equipment side. Therefore, we propose a calibration method for the BS. During the calibration, polynomial function is applied to approximate the nonlinear mismatch factor, and over-the-air training is employed to estimate the polynomial coefficients. After that, the calibration coefficients are computed by maximizing the downlink achievable rate. Simulation results are presented to verify the analytical results and to show the performance of the proposed calibration approach.
\end{abstract}

\section{Introduction}

Massive multiple-input multiple-out (MIMO) has been identified as one of the key enabling technologies for the 5th generation (5G) mobile communication networks \cite{Larsson2014Massive}. With a great number of antennas deployed at the base station (BS) side, massive MIMO can significantly improve the capacity, throughput, and spectral efficiency of wireless communications \cite{Larsson2015Fundamentals}.
As the antenna number increases, acquisition of the channel state information (CSI) becomes a great challenge \cite{Akyildiz20165G}.

To reduce the heavy overhead in obtaining the downlink CSI, massive MIMO systems are typically assumed to operate in time division duplexing (TDD) mode, and BS estimates the downlink CSI from the uplink pilots transmitted by the user equipment (UE) via exploiting the channel reciprocity \cite{Rusek2013Scaling}. In practice, the channel observed by the baseband processor consists of not only the reciprocal wireless propagation channel, but also the radio frequency (RF) gains resulting from the frequency responses of hardware devices, i.e., high power amplifier (HPA), filters, analog-to-digital converter (ADC), and digital-to-analog converter (DAC) \cite{Shan2018a}. Due to the involvement of different devices, the overall gains of the RF chains at BS and UEs are typically asymmetric, which leads to the reciprocity mismatch of the uplink and downlink channels \cite{Shan2017Performance}. 

To study how the reciprocity mismatch affects the system performance, previous works have investigated the impact of the mismatch on the MIMO system with linear precoding techniques, e.g., zero-forcing (ZF) and matched filter (MF). Generally, the reciprocity mismatch can degrade the performances of both ZF and MF precoding techniques \cite{Mi2017Massive}. By comparing the performances of ZF and MF with the mismatch, it can be found that ZF is more sensitive to the reciprocity mismatch and its performance loss is much larger than that of MF, especially in the high signal-to-noise rate (SNR) regime \cite{Zhang2015Large}. As the antenna number of the BS grows, the performances of ZF and MF are asymptotically identical \cite{Raeesi2018Performance}. Wei \emph{et al.} in \cite{Wei2016Impact} analyzed the impact of the reciprocity mismatch at the BS side and the UE side, and showed that both the reciprocity mismatch at the BS side and the UE side degraded the system performance. It is noteworthy that the performance degradation caused by the mismatch at the BS side is much more crucial than that at the UE side. The result reveals that it is important to calibrate the reciprocity mismatch at the BS side. Further, the experimental results in \cite{Jiang2015MIMO} verified the performance loss due to the reciprocity mismatch in the practical precoded system.

As the reciprocity mismatch severely degrades the system performance, reciprocity calibration has attracted great attention in the past decade. Reciprocity calibration can be divided into two types: hardware-circuit calibration and over-the-air calibration \cite{Nie2019A}. The hardware-circuit calibration requires auxiliary components, such as switches and couplers, to connect transmit antennas and receive antennas. Nishimori \emph{et al.} in \cite{Calibration2001Automatic} first proposed an automatic hardware-circuit calibration for the conventional MIMO system. The calibration approach can use the transmit signal to realize a real-time calibration during the BS transmits data streams. Then, the hardware-circuit calibration was applied to the wideband wireless system in \cite{Bourdoux2003Non}, where different subcarriers were independently calibrated. In \cite{Liu2004OFDM}, an auxiliary calibration transceiver structure was proposed for the MIMO system. With the auxiliary transceiver, the calibration loop can be easily extended to the BS with multiple access points, but it inevitably increased the cost of the hardware components. To simplify both the hardware structure and the signal processing, Liu \emph{et al.} proposed a calibration board consisting of switches and attenuators in \cite{JianLiu2006A}. While hardware-circuit calibration works efficiently in conventional MIMO systems with low cost of hardware components, it becomes costly in massive MIMO systems due to the large number of channels to be calibrated. To reduce the cost, a daisy chain interconnection topology of the circuits was applied to hardware-circuit calibration \cite{Benzin2017Internal}, which could reduce the transceiver interconnection effort. To further relieve the connection effort and improve the calibration performance, an optimal interconnection of the hardware-circuit calibration was presented in \cite{Luo2019Massive} by minimizing the Cramer-Rao lower bound in the calibration coefficients.

Unlike the hardware-circuit calibration, the over-the-air calibration only needs to gather the training signals among uncalibrated antennas. An over-the-air calibration called relative calibration was first proposed to calibrate the TDD single-input single-output system in the frequency domain \cite{Guillaud2005A}. Then, Kaltenberger \emph{et al.} proposed an over-the-air calibration for the multi-user MIMO system in \cite{Kaltenberger2010Relative}, which involved both BS and UEs, and they applied the total least square (LS) approach to compute the calibration coefficients. For the wideband system, the frequency-domain calibration has to be applied in each subcarrier, which means that the overhead and complexity of the frequency-domain calibration increases with the number of subcarriers. To reduce the calibration overhead and complexity for the wideband system, a time-domain reciprocity calibration was presented in \cite{Kouassi2012Estimation}, since the number of parameters in the time domain is much less than that in the frequency domain. 

While above reciprocity calibration methods designed for conventional MIMO systems encounter new challenges in massive MIMO systems, due to the heavy overhead of feeding back the CSI from UEs. Based on theoretical and experimental results that the impact of the reciprocity mismatch at the UE side is negligible compared with the reciprocity mismatch at the BS side \cite{r1092359hardware}, several calibration approaches called ``single-side" or ``one-side" calibration were presented for the massive MIMO system, which only calibrate the antennas at the BS. In \cite{Shepard2012Argos}, a single-side calibration method was presented for the massive MIMO Argos prototype. The performance of the Argos calibration is sensitive to the fading channel and relies on the location of the reference antenna. To overcome the shortages of Argos, H. Wei \emph{et al.} in \cite{Wei2016Mutual} presented the mutual coupling calibration, which utilized the strong mutual coupling effect among adjacent antennas rather than the fading channel. To compute the calibration coefficients more efficiently, the LS method was applied to compute the coefficients in \cite{Rogalin2013Hardware}. In \cite{Jiang2018A}, an over-the-air calibration framework was proposed based on some existing calibration schemes. As for distributed massive MIMO systems, a two-stage calibration approach was proposed in \cite{Nie2020Nie2020Relaying} to reduce the overhead of the CSI feedback among remote access points.

All the above mentioned works have simply treated the reciprocity mismatch as a constant. However, as the reciprocity mismatch arises from the asymmetry of the transmit and receive RF chains, and practical RF chains are generally composed of nonlinear components, e.g., the nonlinear HPA \cite{Guerreiro2018Analytical}, the reciprocity mismatch is also nonlinear, which is called nonlinear reciprocity mismatch. With the expected use of millimeter waveband and low-cost RF devices in massive MIMO, nonlinearity will become more severe for 5G and beyond networks. Hence, the nonlinear mismatch needs to be studied in massive MIMO systems. Compared with the linear reciprocity calibration, the nonlinear calibration has three great challenges. As the reciprocity mismatch factor varies with the transmit power, a simple training scheme with the single-power pilot is no longer applicable. Due to the complex expression of the transform characteristic of the nonlinear components, it is difficult to determine the relationship function between the mismatch factor and the transmit power. Further, the relationships between the calibration coefficients are also nonlinear functions so that it is difficult to solve the calibration coefficients.

The nonlinearity compensation of transmitters, especially by the nonlinear predistortion, has been extensively studied in wireless systems \cite{Balti2017Impact,Yu2019Full,Liu2019Linearization}. Previous works on predistortion only focused on offsetting the defects of the transmit RF chain. Since the reciprocity mismatch is jointly caused by the defects of the transmit chain and the receive chain, existing solutions to the nonlinear transmitter compensation techniques are not sufficient to address the nonlinear reciprocal mismatch in massive MIMO systems.


Motivated by the above observations, we investigate the nonlinear reciprocity mismatch in TDD multi-user massive MIMO systems. To study the impact of the nonlinear reciprocity mismatch, we first derive the closed-form expression of the ergodic achievable rate. Then, the performance loss due to the mismatch at the BS side and UE side is analyzed, respectively. Based on these analytical results, we propose a novel nonlinear reciprocity calibration approach. To sample the nonlinear response of the transmit RF chain along with the transmit power, multi-power training pilots are employed in the calibration. Then, polynomial functions are applied to characterize the relationship between the mismatch factor and the transmit power. The polynomial coefficients are estimated by the over-the-air training approach. Finally, to compute the nonlinear calibration coefficients efficiently, we formulate an optimization problem which seeks to maximize the downlink achievable rate. The main contributions of the work can be summarized as follows.
\begin{itemize}
	\item \textbf{Impact analysis of the nonlinear reciprocity mismatch:} Under the nonlinear reciprocity mismatch, we first derive the closed-form expression of the downlink ergodic achievable rate for the multi-user massive MIMO system with ZF precoding. Based on this, the impact of the nonlinear reciprocity mismatch on the system performance is examined.
	\item \textbf{Determine the nonlinear mismatch factor:} To estimate the nonlinear response of the transmit RF chain along with the transmit power, we propose the multi-power training pilots. Then, a polynomial fitting approach is applied to approximate the nonlinear function between the nonlinear reciprocity mismatch factor and the transmit power. After that, the over-the-air training approach is applied to estimate the polynomial coefficients.
	\item \textbf{Toward optimal nonlinear calibration coefficients:} To compute the nonlinear calibration coefficients efficiently, we formulate an optimization problem seeking to maximize the downlink achievable rate. The problem can be transformed into a convex optimization which can be efficiently solved by a fast algorithm.
\end{itemize}

The rest of the paper is organized as follows. Section II describes the system model. The impact of the nonlinear reciprocity mismatch is analyzed in section III. In section IV, nonlinear reciprocity calibration is proposed, including the multi-power training pilots, the nonlinear reciprocity mismatch polynomial fitting, and the optimal nonlinear calibration coefficients. Simulation and numerical results are given in Section V, and the conclusion is given in VI.

Throughout the paper, vectors and matrices are denoted in bold lowercase and uppercase respectively: $ \mathbf{a} $ and $ \mathbf{A} $. Let $ \mathbf{A}^{T} $, $ \mathbf{A}^{H} $, and $ \mathbf{A}^{-1} $ denote the transpose, conjugate transpose, and inverse of a matrix $ \mathbf{A} $ respectively. $ \mathrm{tr}(\cdot) $ stands for the trace operator and $ \mathbb{E}(\cdot) $ represents the expectation operation. Let $ |a| $ denote the amplitude of the complex number $ a $. $ \mathrm{diag}(a_1,\cdots,a_N) $ denotes a $ N $ by $ N $ diagonal matrix with diagonal entries given by $ a_1,\cdots,a_N $. $ \mathcal{N}(\mu,\sigma^2) $ and $ \mathcal{CN}(\mu,\sigma^2) $ represent for normal distribution and complex normal distribution with mean $ \mu $ and variance $ \sigma^2 $, respectively. $\mathcal{U}(a,b)$ denotes uniform distribution on the interval $[a,b]$. $\mathbb{C}$ and $\mathbb{R}$ stand for the complex numbers and real numbers, respectively. Let $[1:N]$ denote the set $\left\lbrace1,2,\cdots,N\right\rbrace$.

\section{System Model}


\begin{figure}
	\centering
	\includegraphics[width=0.6\linewidth]{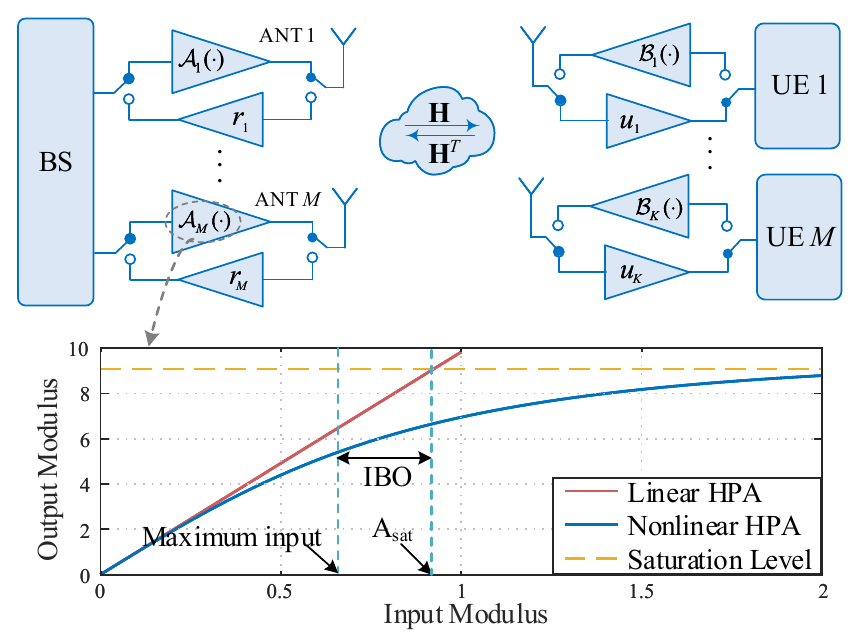}
	\caption{A multi-user massive MIMO system with reciprocity mismatch.}
	\label{fig:ModelHPA}
\end{figure}

We consider a multi-user massive MIMO system consisting of $K$ single-antenna UEs and a BS equipped with $ M $ antennas, as illustrated in Fig. \ref{fig:ModelHPA}, where $M$ and $K$ are large. The transmit RF chains are subject to a typical nonlinear device, i.e., nonlinear HPA. Without loss of generality, a memoryless HPA model called Solid State Power Amplifier (SSPA) is considered in the system, which has been extensively used to characterize the HPA \cite{Balti2017Impact}.

Let $x_{\mathrm{u},k}$ be the transmit signal of the $k$-th UE and $x_{\mathrm{b},m}$ be the transmit signal at the $m$-th antenna of the BS. The transform functions of the HPAs can be characterized as
\begin{equation}
	\hat{x}_{\mathrm{u},k}=\frac{\sqrt{b_0}v_kx_{\mathrm{u},k}}{\sqrt[2v]{1+\left(\frac{|x_{\mathrm{u},k}|}{B_{\mathrm{sat},k}}\right)^{2v}}}=\sqrt{b_0}\mathcal{B}_k(|x_{\mathrm{u},k}|)x_{\mathrm{u},k},
\end{equation}
\begin{equation}
	\hat{x}_{\mathrm{b},m}=\frac{\sqrt{a_0}t_mx_{\mathrm{b},m}}{\sqrt[2v]{1+\left(\frac{|x_{\mathrm{b},m}|}{A_{\mathrm{sat},m}}\right)^{2v}}}=\sqrt{a_0}\mathcal{A}_m(|x_{\mathrm{b},m}|)x_{\mathrm{b},m},
	\label{eq:hpaforbs}
\end{equation}
where $a_0\in\mathbb{R}$ denotes the small-signal amplification gain of the HPAs at the BS, $A_{\mathrm{sat},m}=A_{\mathrm{sat}}a_m\in\mathbb{R}$ is the saturation level of the $m$-th HPA at the BS, $a_m\in\mathbb{R}$ represents the various saturation of different HPAs, $t_m\in\mathbb{C}$ is the vibration of small-signal gain $a_0$, $b_0\in\mathbb{R}$ denotes the small-signal amplification gain of the HPAs at UEs, $B_{\mathrm{sat},k}\in\mathbb{R}$ is the saturation level of HPAs of UEs, $v_k\in\mathbb{C}$ represents the vibration of the small-signal gain $b_0$. Both $a_m$ and the amplitude of $t_m$ follow the log-normal distribution as $\ln a_m\sim \mathcal{N}(0,\delta_{\mathrm{a}}^2)$, $\ln |t_m|\sim \mathcal{N}(0,\delta_{\mathrm{t}}^2)$, and the phase of $t_m$ obeys the uniform distribution as $\angle t_m\sim\mathcal{U}(-\theta_{\mathrm{t}},\theta_{\mathrm{t}})$. Moreover, we use input back-off (IBO) to measure the relationship between the input power and the saturation level of the HPA. From \cite{Balti2017Impact}, IBO is defined as 
\begin{equation}
	\mathrm{IBO}=10\log_{10}\left(\frac{A_{\mathrm{sat}}}{\sigma_{\mathrm{x}}}\right),
\end{equation}
where $\sigma_{\mathrm{x}}^2$ is the average power of the input signal.

The overall channel observed by the baseband processor is composed of the reciprocal wireless propagation channel and the non-reciprocal RF gain. Let $\mathbf{H}=\mathbf{\Phi}^{1/2}\mathbf{H}_{\mathrm{r}}\in\mathbb{C}^{k\times M}$ denote the wireless propagation channel, where $\mathbf{H}_{\mathrm{r}}\in\mathbb{C}^{k\times M}$ is the Rayleigh fading channel with each entry following $\mathcal{CN}(0,1)$, $\mathbf{\Phi}=\mathrm{diag}(\phi_1,\cdots,\phi_K)\in\mathbb{R}^{K\times K}$, and $\phi_k$ denotes the large-scale path loss between the BS and the $k$-th UE. By considering the defects of the transmit and receive RF chain, the overall uplink and downlink channels can be modeled as
\begin{subequations}
	\begin{align}
		&\mathbf{H}_{\mathrm{UL}}=\mathbf{R}\mathbf{H}^T\boldsymbol{\mathcal{B}},\\
		&\mathbf{H}_{\mathrm{DL}}=\mathbf{U}\mathbf{H}\boldsymbol{\mathcal{A}},
		\label{eq:chennelmode}
	\end{align}
\end{subequations}
where $\mathbf{R}=\mathrm{diag}(r_1,\cdots,r_M)$, $r_m\in\mathbb{C}$ denotes the receive RF gain of the $m$-th antenna of the BS, $\boldsymbol{\mathcal{A}}=\mathrm{diag}(\mathcal{A}_1(|x_{\mathrm{b},1}|),\cdots,\mathcal{A}_M(|x_{\mathrm{b},M}|))$, $\mathbf{U}=\mathrm{diag}(u_1,\cdots,u_K)$, $u_k\in\mathbb{C}$ is the receive RF gain of the $k$-th UE, and $\boldsymbol{\mathcal{B}}=\mathrm{diag}(b_1,\cdots,b_K)$, $b_k=\mathcal{B}_k(|x_{\mathrm{u},k}|)\in\mathbb{C}$. Both the amplitudes of $r_m$ and $u_k$ follow log-normal distribution, i.e., $\ln |r_m|\sim\mathcal{N}(0,\delta_{\mathrm{r}}^2)$, $\ln |u_k|\sim\mathcal{N}(0,\delta_{\mathrm{u}}^2)$, and their phases obey uniform distribution, i.e., $\angle r_m\sim \mathcal{U}(-\theta_{\mathrm{r}},\theta_{\mathrm{r}})$, $\angle u_k\sim\mathcal{U}(-\theta_{\mathrm{u}},\theta_{\mathrm{u}})$.

In the downlink, the BS transmits the precoded signal to UEs. Let $\mathbf{y}=[y_1,\cdots,y_K]^T\in\mathbb{C}^{K\times 1}$ denote the signal vector received by all UEs, where $y_k$ is the received signal at the $k$-th UE. The downlink signal can be represented as
\begin{equation}
	\mathbf{y}=\sqrt{a_0}\mathbf{H}_{\mathrm{DL}}\underbrace{\mathbf{W}\mathbf{s}}_{\mathbf{x}_{\mathrm{b}}}+\mathbf{n},
	\label{eq:downtrans}
\end{equation} 
where $\mathbf{s}\in\mathbb{C}^{K\times 1}$ is the symbol vector with zero mean and variance $\mathbb{E}\left\lbrace\mathbf{s}\mathbf{s}^H\right\rbrace=\rho_t\mathbf{I}_K$, $\rho_{\mathrm{t}}\in \mathbb{R}$ represents the average transmit power of the BS, $\mathbf{W}\in\mathbb{C}^{M\times K}$ denotes the precoding matrix, and $\mathbf{n}$ denotes the additive white Gaussian noise (AWGN) vector with $\mathbf{n}\sim\mathcal{CN}(\mathbf{0},\sigma_{\mathrm{n}}^2\mathbf{I}_K)$. In this paper, we consider the ZF precoding technique, and the precoding matrix is given by
\begin{equation}
	\mathbf{W}=\frac{1}{\sqrt{\beta_{\mathrm{ZF}}}}{\mathbf{H}}_{\mathrm{UL}}^*({\mathbf{H}}_{\mathrm{UL}}^T{\mathbf{H}}_{\mathrm{UL}}^*))^{-1},
\end{equation}
where $\beta_{\mathrm{ZF}}$ denotes the normalization scalar defined as
\begin{equation}
	\beta_{\mathrm{ZF}}=\mathbb{E}\left\lbrace\mathrm{tr}\left[({\mathbf{H}}_{\mathrm{UL}}^T{\mathbf{H}}_{\mathrm{UL}}^*)^{-1}\right] \right\rbrace.
	\label{eq:defbetazf}
\end{equation}
As the overall downlink and uplink channels are not reciprocal, the multi-user massive MIMO system suffers from the reciprocity mismatch. Since the multiplicative RF matrices $\boldsymbol{\mathcal{A}}$ and $\boldsymbol{\mathcal{B}}$ are related to the transmit power, we call it nonlinear reciprocity mismatch.

\section{Performance Analysis of Nonlinear Reciprocity Mismatch}\label{sec:analysis}
In this section, we first introduce the performance analysis of the existing linear reciprocity mismatch. Then, the closed-form expression of the ergodic achievable rate is derived for the massive MIMO system in the presence of the nonlinear reciprocity mismatch. We then analyze the impact of the nonlinear reciprocity mismatch on the performance of the downlink transmission.

\subsection{Performance of Existing Linear Reciprocity Mismatch}
The existing works regard the reciprocity mismatch as a constant \cite{Raeesi2018Performance,Zhang2015Large,Wei2016Impact}, i.e., $\mathcal{A}_{m}(|x_{\mathrm{b},m}|)=t_m$ and $\mathcal{B}_{k}(|x_{\mathrm{u},k}|)=v_k$. They also assume that UEs decode the received signals by exploiting the statistical effective channel. Based on \eqref{eq:downtrans}, the signal received by the $k$-th UE can be further expressed as
\begin{equation}
	\begin{split}
		y_k&=\sqrt{a_0}\underbrace{u_k\mathbf{h}_{k}\mathbf{T}\mathbf{w}_{k}}_{h_{\mathrm{eq},k,k}^{\mathrm{lrm}}}s_k+\sqrt{a_0}\sum_{i\neq k}^{K}\underbrace{u_k\mathbf{h}_{k}\mathbf{T}\mathbf{w}_{i}}_{h_{\mathrm{eq},k,i}^{\mathrm{lrm}}}s_i+n_k\\
		&=\underbrace{\sqrt{a_0}\mathbb{E}\left\lbrace h_{\mathrm{eq},k,k}^{\mathrm{lrm}}\right\rbrace s_k}_{\mathrm{ES}_k^{\mathrm{lrm}}}+\mathrm{SI}_k^{\mathrm{lrm}}+\mathrm{MUI}_k^{\mathrm{lrm}}+n_k
	\end{split}
	\label{eq:sinrcon}
\end{equation}
where $\mathbf{h}_{k}$ is the $k$-th row of $\mathbf{H}$, $\mathbf{w}_{k}$ denotes the $k$-th column of $\mathbf{W}$, $\mathbf{T}=\mathrm{diag}(t_1,\cdots,t_M)$, $\mathrm{ES}_k^{\mathrm{lrm}}$ denotes the effective useful signal received by the UE $k$, $\mathrm{SI}_k^{\mathrm{lrm}}$ and $\mathrm{MUI}_k^{\mathrm{lrm}}$ represent the self-interference caused by the uncertainty of the downlink effective channel and the multiple-user interference signal due to the signal for other UEs with
\begin{align}
	&\mathrm{SI}_k^{\mathrm{lrm}}=\sqrt{a_0}\left(h_{\mathrm{eq},k,k}^{\mathrm{lrm}}-\mathbb{E}\left\lbrace h_{\mathrm{eq},k,k}^{\mathrm{lrm}}\right\rbrace\right) s_k,\\
	&\mathrm{MUI}_k^{\mathrm{lrm}}=\sqrt{a_0}\sum_{i\neq k}^{K}h_{\mathrm{eq},k,i}^{\mathrm{lrm}}s_i.
\end{align}

To study the impact of the reciprocity mismatch, previous works used the ergodic achievable rate to characterize the performance of the massive MIMO system as follows.
\begin{deff}[Downlink downlink ergodic achievable rate]
	According to \cite{Shan2017Performance,Raeesi2018Performance,Zhang2015Large}, the ergodic achievable rate can be defined as
	\begin{equation}
		R_k^{\mathrm{lrm}}=\log\left(1+\mathrm{SINR}_k\right),\ \ \forall k\in[1:K],
		\label{eq:ratecon}
	\end{equation}
	where $\mathrm{SINR}_k$ denotes the signal-to-interference-and-noise ratio (SINR) at the $k$-th UE as
	\begin{equation}
		\mathrm{SINR}_k=\frac{\left|\mathbb{E}\left\lbrace h_{\mathrm{eq},k,k}^{\mathrm{lrm}}\right\rbrace\right|^2}{\mathrm{Var}\left\lbrace h_{\mathrm{eq},k,k}^{\mathrm{lrm}}\right\rbrace+\sum_{i\neq k}^{K}\mathbb{E}\left\lbrace h_{\mathrm{eq},k,i}^{\mathrm{lrm}}\right\rbrace+\frac{\sigma_{\mathrm{n}}^2}{a_0\rho_{\mathrm{t}}}}.
	\end{equation}
\end{deff}

Then, the closed-form expression of $\mathrm{SINR}_k$ for the massive MIMO system can be denoted as follows.
\begin{lem}[Closed-form expression of SINR]\label{lem:existSNR}
	According to \cite{Shan2017Performance,Raeesi2018Performance}, the downlink SINR for ZF with the linear reciprocity mismatch can be given by
	\begin{equation}
		\mathrm{SINR}_{\mathrm{ZF},k}=\frac{a_0\rho_{\mathrm{t}}\frac{M-K}{\mathrm{tr}\left\lbrace \mathbf{\Phi}^{-2}\right\rbrace}\mathrm{sinc}^2(\theta_{\mathrm{t}})\mathrm{sinc}^2(\theta_{\mathrm{r}})}{e^{\delta_{\mathrm{r}}^2+\delta_{\mathrm{v}}^2-\delta_{\mathrm{t}}^2}\left(a_0\rho_{\mathrm{t}}\phi_k^2\frac{M-K}{M}\varepsilon_1+\sigma_{\mathrm{n}}^2\right)},
	\end{equation}
	where $\varepsilon_1=e^{2\delta_{\mathrm{t}}^2}+e^{2\delta_{\mathrm{r}}^2}-2\mathrm{sinc}(\theta_{\mathrm{t}})\mathrm{sinc}(\theta_{\mathrm{r}})e^{(\delta_{\mathrm{t}}^2-\delta_{\mathrm{r}}^2)/{2}}$.
\end{lem}
From Lemma \ref{lem:existSNR}, it can be seen that the reciprocity mismatch leads interference and degrades the SINR. Although Lemma \ref{lem:existSNR} shows the impact of the linear reciprocity mismatch, the nonlinear reciprocity mismatch has never been studied. Therefore, in the remainder of this section, we focus on the impact of the nonlinear reciprocity mismatch.

\subsection{Downlink SINDR and Ergodic Achievable Rate}
TO derive the closed-form expression of the achievable rate, we apply the Bussgang's theory \cite{Balti2017Impact,Rowe1982Memoryless} to characterize the nonlinearities of the HPAs at the BS side. The theory states that the output of the nonlinear HPA can be expressed in terms of a linear scale parameter of the input signal and a nonlinear distortion which obeys the complex circular Gaussian distribution and is independent of the input. At the BS side, the output signal of HPAs can be further denoted as follows.
\begin{lem}[{\cite[Eq. (18)]{Balti2017Impact}}]\label{lem:bussgangtheo}
	By employing the Bussgang's linearization theory, the transform characteristics of the SSPA can be further characterized as
	\begin{equation} 
		\hat{x}_{\mathrm{b},m}=g_mx_{\mathrm{b},m}+d_m,\ \forall m\in[1:M],
		\label{eq:bussgang}
	\end{equation}
	where $g_m$ is the linear scale parameter denoted as
	\begin{equation}
		g_m=\frac{\mathbb{E}\left\lbrace x_{\mathrm{b},m}^*\hat{x}_{\mathrm{b},m}\right\rbrace}{\mathbb{E}\left\lbrace|x_{\mathrm{b},m}|^2\right\rbrace}=t_m\mu\left(\frac{A_{\mathrm{sat},m}}{\sigma_{x,m}}\right),
		\label{eq:gengm}
	\end{equation} $\mu(x)=\frac{x}{2}\left[2x-\sqrt{\pi}\mathrm{erfc}(x)\mathrm{exp}(x^2)(2x^2-1)\right]$,
	$d_m\in\mathbb{C}$ denotes the nonlinear Gaussian distortion with zero mean and variance $\sigma_{\mathrm{d},m}^2$ as
	\begin{equation}
		\sigma_{\mathrm{d},m}^2=\mathbb{E}\left\lbrace|\hat{x}_{\mathrm{b},m}|^2\right\rbrace-g_m\mathbb{E}\left\lbrace \hat{x}_{\mathrm{b},m}^*x_{\mathrm{b},m}\right\rbrace=|t_m|^2\lambda_m\left(\sigma_{x,m}\right)
		\label{eq:sigmad}
	\end{equation}
	with
	\begin{equation}
		\begin{split}
			\lambda_m(x)={A_{\mathrm{sat},m}^2}+\frac{A_{\mathrm{sat},m}^4}{x^2}\mathrm{exp}\left(\frac{A_{\mathrm{sat},m}^2}{x^2}\right)\mathrm{Ei}\left(-\frac{A_{\mathrm{sat},m}^2}{x^2}\right)
			-{x^2}\mu\left(\frac{A_{\mathrm{sat},m}}{x}\right)^2,
		\end{split} 
	\end{equation}
	and $\mathrm{Ei}(x)=\int_{-\infty}^{x}t^{-1}e^t\mathrm{d}t\ (x< 0)$ is an exponential integral function.
\end{lem}

According to Lemma \ref{lem:bussgangtheo}, the downlink signal received by UEs can be rewritten as
\begin{equation}
	\begin{split}
		\mathbf{y}=\sqrt{a_0}\mathbf{UH}\hat{\mathbf{x}}_{\mathrm{b}}+\mathbf{n}=\sqrt{a_0}\mathbf{UHG}\mathbf{x}_{\mathrm{b}}+\sqrt{a_0}\mathbf{UH}\mathbf{d}+\mathbf{n},
	\end{split}
\end{equation}
where $\mathbf{G}=\mathrm{diag}(g_1,\cdots,g_M)$ and $\mathbf{d}=[d_1,\cdots,d_M]^T$.

Similar to \eqref{eq:sinrcon}, we assume that the UEs decode the signals by exploiting the statistical effective channel gain, and the signal received by the $k$-th UE can be further denoted as
\begin{equation}
	\begin{split}
		y_k=\underbrace{\sqrt{a_0}\mathbb{E}\left\lbrace h_{\mathrm{eq},k,k}\right\rbrace s_k}_{\mathrm{ES}_k}+\mathrm{SI}_k+\mathrm{MUI}_k+\mathrm{NLD}_k+n_k,
	\end{split}
	\label{eq:dlsignal}
\end{equation}
where $h_{\mathrm{eq},k,k}=u_k\mathbf{h}_k\mathbf{Gw}_k$ denotes the $k$-th row of $\mathbf{H}_{\mathrm{eq}}=\mathbf{UHGW}$, $\mathrm{ES}_k$ denotes the effective useful signal received by the UE $k$, $\mathrm{SI}_k$ and $\mathrm{MUI}_k$ represent the self-interference and the multi-user interference expressed as
\begin{align}
	&\mathrm{SI}_k=\sqrt{a_0}\left(h_{\mathrm{eq},k,k}-\mathbb{E}\left\lbrace h_{\mathrm{eq},k,k}\right\rbrace\right) s_k,\\
	&\mathrm{MUI}_k=\sqrt{a_0}\sum_{i\neq k}^{K}h_{\mathrm{eq},k,i}s_i,
\end{align}
and $\mathrm{NLD}_k$ denotes the nonlinear distortion for the $k$-th UE due to the nonlinear HPAs of the BS
\begin{equation}
	\mathrm{NLD}_k=a_0u_k\mathbf{h}_k\mathbf{d}.
\end{equation}

Compared with the linear reciprocity mismatch, the signal in the presence of the nonlinear reciprocity mismatch has an extra nonlinear distortion term $\mathrm{NLD}_k$. Therefore, we use the signal-to-interference-plus-noise-and-distortion ratio (SINDR) to measure the ratio between the average power of the useful signal and the sum average power of the interference, noise, and distortion.
\begin{deff}[Downlink SINDR and achievable rate]\label{def:SINDRandrate}
	Based on \eqref{eq:dlsignal}, the SINDR at the $k$-th UE can be denoted as 
	\begin{equation}
		\gamma_k=\frac{a_0\rho_{\mathrm{t}}|\mathbb{E}\left\lbrace h_{\mathrm{eq},k,k}\right\rbrace|^2}{a_0\rho_{\mathrm{t}}\mathrm{Var}\left\lbrace h_{\mathrm{eq},k,k}\right\rbrace+a_0\rho_{\mathrm{t}}\sum_{i\neq k}^{K}\mathbb{E}\left\lbrace|h_{\mathrm{eq},k,i}|^2\right\rbrace+a_0|u_k|^2\mathbb{E}\left\lbrace\|\mathbf{h}_k\boldsymbol{\Sigma}_{\mathrm{d}}^{\frac{1}{2}}\|^2\right\rbrace+\sigma_n^2},
		\label{eq:sinrdef}
	\end{equation}
	where $\boldsymbol{\Sigma}_{\mathrm{d}}=\mathrm{diag}(\sigma_{\mathrm{d},1}^2,\cdots,\sigma_{\mathrm{d},M}^2)$ is the variance matrix of the distortion. The downlink ergodic achievable rate in the presence of the nonlinear reciprocity mismatch can be defined as
	\stepcounter{equation}
	\begin{equation}
		R_k=\log\left(1+\gamma_k\right),\ \ \forall k\in[1:K].
		\label{eq:ratedef}
	\end{equation}
\end{deff}

As UEs treat the sum of the noise and the interference as equivalent AWGN uncorrelated with the useful signal, the ergodic rate in Definition \ref{def:SINDRandrate} is upper bounded by the capacity and legitimately achievable \cite{Shan2017Performance,Zhang2015Large,Raeesi2018Performance}. Then, we can derive the closed-form expression of the SINDR for the multi-user massive MIMO system with ZF in the presence of the nonlinear reciprocity mismatch.
\begin{prop}[SINDR with nonlinear reciprocity mismatch]\label{theo:ratezf}
	For a multi-user massive MIMO system with ZF, the closed-form expression of the SINDR in the presence of the nonlinear HPA and the reciprocity mismatch can be denoted as
	\begin{equation}
		\gamma_{\mathrm{ZF},k}=\frac{\varUpsilon_{\mathrm{ZF},k}^{\mathrm{ES}}}{\varUpsilon_{\mathrm{ZF},k}^{\mathrm{SI}}+\varUpsilon_{\mathrm{ZF},k}^{\mathrm{MUI}}+\varUpsilon_{\mathrm{ZF},k}^{\mathrm{NLD}}+\sigma_n^2},
		\label{eq:sinrzf}
	\end{equation}
	where $	\varUpsilon_{\mathrm{ZF},k}^{\mathrm{ES}}$ denotes the power of the effective signal given by
	\begin{equation}
		\varUpsilon_{\mathrm{ZF},k}^{\mathrm{ES}}=\frac{a_0\rho_{\mathrm{t}}(M-K)|u_k\mathrm{tr}\left\lbrace\mathbf{G_{\mathrm{ZF}}R}^*\right\rbrace|^2}{M|b_k|^2\mathrm{tr}\left\lbrace\boldsymbol{(\mathcal{B}\Phi^2\mathcal{B}^*)^{-1}}\right\rbrace\mathrm{tr}\left\lbrace\mathbf{RR}^*\right\rbrace}
		\label{eq:poweres}
	\end{equation}
	with $\mathbf{G}_{\mathrm{ZF}}=\mathrm{diag}(g_{\mathrm{ZF},1},\cdots,g_{\mathrm{ZF}})$ and $g_{\mathrm{ZF},m}=t_m\mu\left(A_{\mathrm{Ast},m}\sqrt{\mathrm{tr}\left\lbrace\mathbf{RR}^*\right\rbrace}/(\sqrt{|r_m|^2\rho_{\mathrm{t}}})\right)$. $\varUpsilon_{\mathrm{ZF},k}^{\mathrm{SI}}$ denotes the power of the self-interference and can be given by
	\begin{equation}
		\varUpsilon_{\mathrm{ZF},k}^{\mathrm{SI}}=\frac{a_0\rho_{\mathrm{t}}|u_k|^2\mathrm{tr}\left\lbrace(\mathbf{G}_{\mathrm{ZF}}-\alpha\mathbf{R})(\mathbf{G}_{\mathrm{ZF}}^*-\alpha^*\mathbf{R}^*) \right\rbrace}{(M-K)^{-1}M^2|b_k|^2\mathrm{tr}\left\lbrace(\boldsymbol{\mathcal{B}}\mathbf{\Phi}^2\boldsymbol{\mathcal{B}}^*)^{-1}\right\rbrace}
		\label{eq:powersi}
	\end{equation}
	with $\alpha=\frac{1}{M}\mathrm{tr}\left\lbrace \mathbf{R}^{-1}\mathbf{G}_{\mathrm{ZF}}\right\rbrace$. Further, $\varUpsilon_{\mathrm{ZF},k}^{\mathrm{MUI}}$ is the power of the multi-user interference, which can be denoted as
	\begin{equation}
		\varUpsilon_{\mathrm{ZF},k}^{\mathrm{MUI}}=\sum_{i\neq k}^{K}\frac{a_0\rho_{\mathrm{t}}\phi_k^2\mathrm{tr}\left\lbrace (\mathbf{G}_{\mathrm{ZF}}-\alpha\mathbf{R})(\mathbf{G}_{\mathrm{ZF}}^*-\alpha^*\mathbf{R}^*)\right\rbrace}{(M-K)^{-1}M^2|u_k^{-1}b_i|^2\phi_i^2\mathrm{tr}\left\lbrace(\boldsymbol{\mathcal{B}}\mathbf{\Phi}^2\boldsymbol{\mathcal{B}}^*)^{-1}\right\rbrace}.
		\label{eq:powermui}
	\end{equation}
	Finally, $\varUpsilon_{\mathrm{ZF},k}^{\mathrm{NLD}}$ is the power of the nonlinear distortion given by
	\begin{equation}
		\varUpsilon_{\mathrm{ZF},k}^{\mathrm{NLD}}=a_0|u_k|^2\phi_k^2\mathrm{tr}\left\lbrace\mathbf{\Sigma}_{\mathrm{ZF}}\right\rbrace
		\label{eq:powernld}
	\end{equation}
	with $\mathbf{\Sigma}_{\mathrm{ZF}}=\mathrm{diag}(\sigma_{\mathrm{ZF},1}^2,\cdots,\sigma_{\mathrm{ZF},M}^2)$ and $\sigma_{\mathrm{ZF},m}^2=|t_m|^2\lambda_m\left(\sqrt{{|r_m|^2\rho_{\mathrm{t}}}/{\mathrm{tr}\left\lbrace\mathbf{RR}^*\right\rbrace}}\right)$.
\end{prop}

\begin{IEEEproof}
	See Appendix \ref{append:rateZF}.
\end{IEEEproof}

From Proposition \ref{theo:ratezf}, we find that the self-interference and the multi-user interference are completely determined by the difference of $g_{\mathrm{ZF},m}$ and $\alpha r_m$ from the nonlinear reciprocity mismatch at the BS side. Since $\sigma_{\mathrm{ZF},m}$ increases with the transmit power, the distortion of the nonlinear HPA of the BS decreases the SINDR. These results reveal that the nonlinear reciprocity mismatch in the BS definitely degrades the performance of the ZF-precoded MIMO system. Besides, the impact of the nonlinear reciprocity mismatch at the UE side on the performance relies on the $|u_k|^2/|b_k|^2$ whose impact is difficult to determine. In the following section, we will give more insights into the impact of the nonlinear reciprocity mismatch.

\subsection{Impact Analysis of Nonlinear Reciprocity Mismatch}

To further analyze the impact of the nonlinear reciprocity mismatch, we derive the average achievable rate of the multi-user massive MIMO system. The average achievable rate can be defined as $R=\frac{1}{K}\sum_{k=1}^{K}R_k$, whose closed-form expression can be derived as follows.

\begin{prop}[Average achievable rate]\label{coro:arate}
	Assume that the SINDR $\gamma_{\mathrm{ZF},k}$ large than $1$. Based on the closed-form expression of SINDR denoted in \eqref{eq:sinrzf}, the average ergodic achievable rate of the multi-user massive MIMO system can be denoted as
	\begin{equation}
		R=R_{\mathrm{Ideal}}-\Delta R_{\mathrm{BS}}-\Delta R_{\mathrm{UE}},
	\end{equation}
	where $R_{\mathrm{Ideal}}$ denotes the ideal achievable rate without HPA nonlinearity and reciprocity mismatch given by
	\begin{equation}
		R_{\mathrm{Ideal}}=\log\left(\frac{M-K}{\mathrm{tr}\left\lbrace \mathbf{\Phi}^{-2}\right\rbrace}\cdot\frac{\rho_{\mathrm{t}}a_0}{\sigma_{n}^2}\right).
	\end{equation}
	$\Delta R_{\mathrm{BS}}$ denotes the average performance loss due to the nonlinear reciprocity mismatch at the BS side and is given by
	\begin{equation}
		\Delta R_{\mathrm{BS}}=\frac{1}{K}\sum_{k=1}^{K}\log\left(\frac{\frac{M-K}{M}\rho_{\mathrm{t}}a_0\phi_k^2\varepsilon_2+\sigma_{\mathrm{eq},k}^2}{\frac{1}{M}\sigma_{\mathrm{n}}^2|\mathrm{tr}\left\lbrace \mathbf{G}_{\mathrm{ZF}}\mathbf{R}^*\right\rbrace|^2(\mathrm{tr}\left\lbrace \mathbf{RR}^*\right\rbrace)^{-1}}\right)
		\label{eq:ratebs}
	\end{equation}
	with $ \varepsilon_2=\mathrm{tr}\left\lbrace (\mathbf{G}_{\mathrm{ZF}}-\alpha\mathbf{R})(\mathbf{G}_{\mathrm{ZF}}-\alpha\mathbf{R})^*\right\rbrace/M$, $\sigma_{\mathrm{eq},k}^2=a_0\phi_k^2\mathrm{tr}\left\lbrace\mathbf{\Sigma}_{\mathrm{ZF}}\right\rbrace+\sigma_{\mathrm{n}}^2/|u_k|^2$. Finally, $\Delta R_{\mathrm{UE}}$ denotes the average performance loss caused by the nonlinear reciprocity mismatch at the UE side denoted as
	\begin{equation}
		\Delta R_{\mathrm{UE}}=\frac{1}{K}\sum_{k=1}^{K}\log\left(\frac{|b_k|^2}{K|u_k|^2}\mathrm{tr}\left\lbrace \boldsymbol{(\mathcal{B}\mathcal{B}^*)^{-1}}\right\rbrace\right).
		\label{eq:ratelossUE}
	\end{equation}
	
\end{prop}

\begin{IEEEproof}
	By substituting \eqref{eq:poweres}-\eqref{eq:powernld} into \eqref{eq:sinrzf}, the complete expression of SINDR can be given by
	\begin{equation}
		\begin{split}
			\gamma_{\mathrm{ZF}}=\frac{(M-K)\rho_{\mathrm{t}}a_0|u_k\mathrm{tr}\left\lbrace \mathbf{G}_{\mathrm{ZF}}\mathbf{R}^*\right\rbrace|^2}{M|b_k|^2\mathrm{tr}\left\lbrace \mathbf{RR}^*\right\rbrace\mathrm{tr}\left\lbrace \boldsymbol{(\mathcal{B}\mathbf{\Phi}^2\mathcal{B}^*)^{-1}}\right\rbrace}\cdot
			\left(
			\frac{M-K}{M}\rho_{\mathrm{t}}a_0|u_k|^2\phi_k^2\varepsilon_2+\sigma_{\mathrm{eq,u}}^2\right)^{-1},
		\end{split}
	\end{equation}
	where $\varepsilon_2=\mathrm{tr}\left\lbrace (\mathbf{G}_{\mathrm{ZF}}-\alpha\mathbf{R})(\mathbf{G}_{\mathrm{ZF}}-\alpha\mathbf{R})^*\right\rbrace/M$, and $\sigma_{\mathrm{eq,u}}^2=a_0\phi_k^2|u_k|^2\mathrm{tr}\left\lbrace\mathbf{\Sigma}_{\mathrm{ZF}}\right\rbrace+\sigma_{\mathrm{n}}^2$. Then, assuming SINDR larger than $1$, the average ergodic achievable rate can be further denoted as
	\begin{equation}
		\begin{split}
			R&\approx \frac{1}{K}\sum_{k=1}^{K}\log\left(\gamma_{\mathrm{ZF},k}\right)\\
			&\overset{\equlabel}{=}\underbrace{\log\left(\frac{M-K}{\mathrm{tr}\left\lbrace \mathbf{\Phi}^{-2}\right\rbrace}\frac{\rho_{\mathrm{t}}a_0}{\sigma_{\mathrm{n}}^2}\right)}_{R_{\mathrm{Ideal}}}-\underbrace{\frac{1}{K}\sum_{k=1}^{K}\log\left(\frac{\mathrm{tr}\left\lbrace \boldsymbol{(\mathcal{B}\mathcal{B}^*)^{-1}}\right\rbrace}{K|u_k|^2|b_k|^{-2}}\right)}_{\Delta R_{\mathrm{UE}}}\\
			&\quad -\underbrace{\frac{1}{K}\sum_{k=1}^{K}\log\left(\frac{\frac{M-K}{M}\rho_{\mathrm{t}}a_0\phi_k^2|u_k|^2\varepsilon_2+\sigma_{\mathrm{eq,u}}^2}{\frac{1}{M}\sigma_{\mathrm{n}}^2|\mathrm{tr}\left\lbrace \mathbf{G}_{\mathrm{ZF}}\mathbf{R}^*\right\rbrace|^2(\mathrm{tr}\left\lbrace \mathbf{RR}^*\right\rbrace)^{-1}}\right)}_{\Delta R_{\mathrm{BS}}},
		\end{split}
	\end{equation}
	where $(a)$ holds due to $\frac{1}{K}\mathrm{tr}\left\lbrace \boldsymbol{(\mathcal{B}\mathbf{\Phi}^2\mathcal{B}^*)^{-1}}\right\rbrace-\frac{1}{K}\mathrm{tr}\left\lbrace \boldsymbol{(\mathcal{B}\mathcal{B}^*)^{-1}}\right\rbrace\frac{1}{K}\mathrm{tr}\left\lbrace \mathbf{\Phi}^{-2}\right\rbrace\overset{K\rightarrow\infty}{\longrightarrow}0$ \cite[Lemma 9]{KrishnanLinear2016}.
	By using the law of large number (LLN), $\Delta R_{\mathrm{BS}}$ can be further denoted as
	\begin{equation}
		\Delta R_{\mathrm{BS}}=\frac{1}{K}\sum_{k=1}^{K}\log\left(\frac{\frac{M-K}{M}\rho_{\mathrm{t}}a_0\phi_k^2\varepsilon_2+\sigma_{\mathrm{eq},k}^2}{\frac{1}{M}\sigma_{\mathrm{n}}^2|\mathrm{tr}\left\lbrace \mathbf{G}_{\mathrm{ZF}}\mathbf{R}^*\right\rbrace|^2(\mathrm{tr}\left\lbrace \mathbf{RR}^*\right\rbrace)^{-1}}\right),
	\end{equation}
	where $\sigma_{\mathrm{eq},k}^2=a_0\phi_k^2\mathrm{tr}\left\lbrace\mathbf{\Sigma}_{\mathrm{ZF}}\right\rbrace+\sigma_{\mathrm{n}}^2/|u_k|^2$. Therefore, Proposition \ref{coro:arate} holds.
\end{IEEEproof}

From \eqref{eq:ratelossUE}, the average ergodic achievable rate degradation caused by the nonlinear reciprocity mismatch at the UE side can be further denoted as
\begin{equation}
	\begin{split}
		\Delta R_{\mathrm{UE}}&=\frac{1}{K}\sum_{k=1}^{K}\log\left(\frac{|b_k|^2}{K|u_k|^2}\sum_{q=1}^{K}|b_{q}|^{-2}\right)\\
		&=\log\left(\frac{1}{K}\sum_{k=1}^{K}\frac{1}{|b_k|^2}\right)-\frac{1}{K}\sum_{k=1}^{K}\log\left(\frac{1}{|b_k|^2}\right)-\frac{1}{K}\sum_{k=1}^{K}\log(|u_k|^{2})\\
		&\overset{\equlabel}{=} \log\left(\frac{1}{K}\sum_{k=1}^{K}\frac{1}{|b_k|^2}\right)-\frac{1}{K}\sum_{k=1}^{K}\log\left(\frac{1}{|b_k|^2}\right)
		\overset{\equlabel}{\geq}0,
	\end{split}
	\label{eq:impactue}
\end{equation}
where $(b)$ holds due to LLN, $(c)$ is conditioned on the Jensen's inequality as $\log(\frac{1}{K}\sum_{k=1}^{K}x_k)\geq \frac{1}{K}\sum_{k=1}^{K}\log(x_k)$, and the equality holds if and only if $x_1=\cdots=x_K$.

\begin{rem}[Impact of the nonlinear reciprocity mismatch at the UE side]
	From \eqref{eq:impactue}, $\Delta R_{\mathrm{UE}}$ is always larger than $0$, which reveals that the nonlinear reciprocity mismatch at the UE side always degrades the performance of the multi-user massive MIMO system. Further, the performance loss caused by the nonlinear mismatch at the UE side is not related to the downlink transmit power.
\end{rem}

According to \eqref{eq:ratebs}, the performance degradation caused by the nonlinear reciprocity mismatch at the BS side increases with the average transmit power $\rho_{\mathrm{t}}$. Due to the complex expression of $g_{\mathrm{ZF},m}$ and $\sigma_{\mathrm{ZF},m}$, it is difficult to thoroughly analyze the impact of the average power and the saturation level $A_{\mathrm{sat},m}$ from \eqref{eq:ratebs}. In practice, the power of the input signal of the HPA is less than a maximum acceptable input power, and the IBO is larger than a threshold. Therefore, to further analyze the impact of the nonlinear amplification and the reciprocity mismatch at the BS side, we consider a special case where the IBO is larger than zero and the hardware of UEs is perfect. Based on the assumption, the SINDR can be further denoted as follows.

\begin{prop}[SINDR with large IBO]\label{coro:sinrlimitinterzf}
	Suppose that the hardware and HPAs of UEs are perfect, i.e., $u_k=v_k=1$ and $B_{\mathrm{sat}}\gg |x_{\mathrm{u},k}|$. When the saturation level of the HPA is larger than the average power of the input signal of the BS, the downlink SINDR of the multi-user massive MIMO system with ZF in the presence of the nonlinear reciprocity mismatch can be further given by
	\begin{equation}
		\gamma_{\mathrm{ZF},k}^{\mathrm{LIBO}}=\frac{a_0\varepsilon_4\frac{(M-K)}{\mathrm{tr}\left\lbrace \mathbf{\Phi}^{-2}\right\rbrace}\left(1-\frac{2\rho_{\mathrm{t}}e^{\delta_{\mathrm{a}}^2}}{MA_{\mathrm{sat}}^2}\right)\rho_{\mathrm{t}}} {\frac{M-K}{M}a_0\phi_k^2\varepsilon_3\left(1-\frac{2\rho_{\mathrm{t}}e^{\delta_{\mathrm{a}}^2}}{MA_{\mathrm{sat}}^2}\right)\rho_{\mathrm{t}}+\sigma_{\mathrm{n}}^2},
		\label{eq:highpowerZF}
	\end{equation}
	where $\varepsilon_3=e^{2\delta_t^2}+(e^{2\delta_r^2}-2)e^{\delta_t^2+\delta_r^2}\mathrm{sinc}^2(\theta_{\mathrm{t}})\mathrm{sinc}^2(\theta_{\mathrm{r}})$, and $\varepsilon_4=\mathrm{sinc}^2(\theta_{\mathrm{t}})\mathrm{sinc}^2(\theta_{\mathrm{r}})e^{\delta_t^2-\delta_r^2}$.
\end{prop}

\begin{IEEEproof}
	Proof see Appendix \ref{proof:sinrlimitinterzf}.
\end{IEEEproof}

From \eqref{eq:highpowerZF}, since the average transmit power $\rho_{\mathrm{t}}/M$ is less than $A_{\mathrm{sat}}^2$, the SINDR always increases with $\rho_{\mathrm{t}}$. In addition, the SINDR also increases with the saturation level $A_{\mathrm{sat}}$. When the sum power of the self-interference and multi-user interference dominates the denominator of SINDR, SINDR would approach an upper limit subject to the ratio of $\varepsilon_3$ and $\varepsilon_4$. Moreover, increasing the number of the antennas  $M$ at the BS can also increase SINDR, which can be explained from two aspects. The antenna array gain of BS increases as more antennas are deployed at the BS. Increasing the number of antennas means that the average power allocated to each antenna decreases, which reduces the nonlinearity of the HPA.

\begin{rem}[Impact of the nonlinear reciprocity at the BS side]
	The reciprocity mismatch at the BS side generates the self-interference and multi-user interference, which dramatically degrades the system performance. Besides, the performance loss due to the mismatch at the BS side increases with the average transmit power. Further, the nonlinearity of the HPA can exacerbate the reciprocity mismatch and the performance loss.
\end{rem}

From the above analytical results, the performance loss due to the mismatch at the BS side increases with the transmit power, while the performance loss due to the mismatch at the UE side remains constant. This shows that the impact of the nonlinear reciprocity mismatch at the BS side is much more severe than that at the UE side. Therefore, the reciprocity calibration at the BS side is essential for the multi-user massive MIMO system to improve the system performance.

\section{Calibration of Nonlinear Reciprocity Mismatch}
Based on the analytical results in the above section, the nonlinear reciprocity mismatch causes dramatic performance degradation. In this section, we propose a nonlinear reciprocity calibration approach for the BS to mitigate the performance loss due to the nonlinear reciprocity mismatch. 

\subsection{Existing Work on Linear Reciprocity Calibration}
The reciprocity calibration aims at making the ratio of downlink and uplink channel equal, i.e.,
\begin{equation}
	c_m\frac{h_{\mathrm{DL},m,k}}{h_{\mathrm{UL},m,k}}=c_i\frac{h_{\mathrm{DL},i,k}}{h_{\mathrm{UL}.i,k}},\ \forall m,i\in[1:M],\ k\in[1:K],
	\label{eq:calmis}
\end{equation}
where $c_m$ is the calibration coefficient of the $m$-th antenna at the BS. 

In the massive MIMO system with the linear reciprocity mismatch, ${h_{\mathrm{DL},m,k}}/{h_{\mathrm{UL},m,k}}=t_m/r_m$. Hence, \eqref{eq:calmis} can be further denoted as
\begin{equation}
	c_m\frac{t_{m}}{r_m}=c_i\frac{t_i}{r_i},\ \forall m,i\in[1:M].
	\label{eq:caequprimal}
\end{equation}
From \eqref{eq:caequprimal}, it is clear that the reciprocity calibration requires the knowledge of the ratio of $t_m$ and $r_m$. In the conventional reciprocity calibration, $f_m={t_{m}}/{r_m}$ $(m\in[1:M])$ can be estimated by using the over-the-air training \cite{Shepard2012Argos,Rogalin2013Hardware,Jiang2018A}. The training pilot sequence of the $m$-th antenna at the BS can be denoted as $\mathbf{x}_m=\left[ x_m[1],\cdots,x_m[Q]\right]^T$ with $\mathbb{E}\left\lbrace|x_{m}[q]|^2\right\rbrace=\rho_{\mathrm{c}}$, where $Q$ is the pilot length. Let $\mathbf{y}_{m,i}$ denote the training signal transmitted by the $m$-th antenna and received by the $i$-th antenna, and let $\mathbf{y}_{i,m}$ represent the opposite direction signal. The received signals can be denoted as
\begin{subequations}
	\begin{align}
		&\mathbf{y}_{m,i}=a_0\omega_{m,i}t_mr_i\mathbf{x}_m+\mathbf{n}_{m,i},\\
		&\mathbf{y}_{i,m}=a_0\omega_{m,i}t_ir_m\mathbf{x}_i+\mathbf{n}_{i,m},
	\end{align}
\end{subequations}
where $\omega_{m,i}$ denotes the wireless propagation channel between the $m$-th antenna and the $i$-th antenna, $\mathbf{n}_{m,i}$ and $\mathbf{n}_{i,m}$ are AWGN vector. According to \cite{Rogalin2013Hardware}, the reciprocity mismatch factors can be computed by
\begin{equation}
	\hat{\mathbf{f}}=[1,-\bar{\mathbf{y}}_1^T\bar{\mathbf{Y}}_2^*(\bar{\mathbf{Y}}_2^T\bar{\mathbf{Y}}_2^*)^{-1}],
\end{equation}
where $\bar{\mathbf{y}}_1$ is the first column of the matrix $\bar{\mathbf{Y}}$, $\bar{\mathbf{Y}}_2$ consist of the second column to the last column of $\bar{\mathbf{Y}}$, and $\bar{\mathbf{Y}}\in \mathbb{C}^{M\times M}$ is defined as
\begin{equation}
	[\bar{\mathbf{Y}}]_{m,i}=\left\lbrace\begin{matrix}
		\sum_{j\neq m}^{M}|\mathbf{x}_{m}^T\mathbf{y}_{j,m}|^2,&\mathrm{when}\ m=i,\\
		-\mathbf{y}_{i,m}^H\mathbf{x}_m^*\mathbf{x}_{i}^T\mathbf{y}_{m,i},&\mathrm{when}\ m\neq i.
	\end{matrix}\right.
\end{equation}
Then, the linear calibration coefficients can be legitimately computed by
\begin{equation}
	\hat{\mathbf{c}}=\left[\frac{c_0}{\hat{f}_1},\cdots,\frac{c_0}{\hat{f}_M}\right]^T,
\end{equation}
where $c_0$ can be any non-zero constant. Finally, during the data transmission, the precoding matrix for implementing the linear reciprocity calibration can be given by $\mathbf{W}_{\mathrm{c}}=\frac{1}{\sqrt{\beta_{\mathrm{c}}}}\mathrm{diag}(\mathbf{c})\mathbf{W}$ where $\beta_{\mathrm{c}}=\mathrm{tr}\left\lbrace \mathrm{diag}(\mathbf{c})\mathbf{W}\mathbf{W}^H\mathrm{diag}(\mathbf{c}^*)\right\rbrace$.

\begin{figure}
	\centering
	\includegraphics[width=0.6\linewidth]{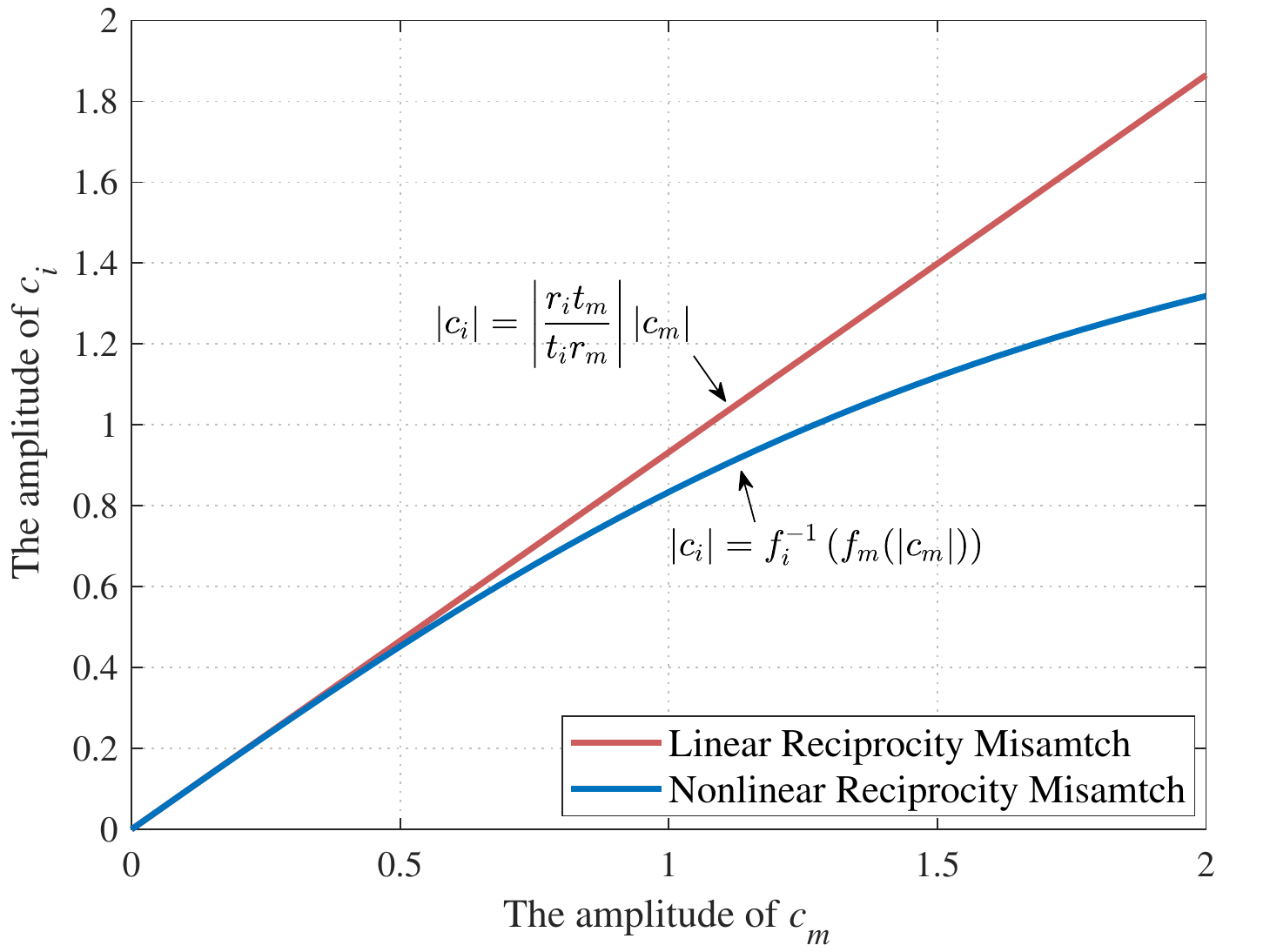}
	\caption{The relationship of the calibration coefficients $c_m$ and $c_i\ (i\neq m)$ $\forall m,i\in[1:M]$. The function $\varphi_i^{-1}(x)$ denotes the inverse function of $x=\varphi_i(|c_i|)=|c_i\mu_{i}(|c_i|\sigma_{\mathrm{x},m})|$, and $\varphi_m(|c_m|)=|c_m\mu_m(|c_m|\sigma_{\mathrm{x},m})|$.}
	\label{eq:calibrationcoef}
\end{figure}

In the massive MIMO system with the nonlinear reciprocity mismatch, ${h_{\mathrm{DL},m,k}}/{h_{\mathrm{UL},m,k}}=\frac{t_m}{r_m}\mu(A_{\mathrm{sat},m}/\sigma_{\mathrm{x},m})=\mu_m(\sigma_{\mathrm{x},m})=\bar{g}_m$, and \eqref{eq:calmis} can be rewritten as
\begin{equation}
	c_m\mu_m(|c_m|\sigma_{\mathrm{x},m})=c_i\mu_i(|c_i|\sigma_{\mathrm{x},i}),\ \forall m,i\in[1:M].
	\label{eq:caequprimalnl}
\end{equation}
From the equation, the nonlinear reciprocity calibration requires the knowledge of the function $\bar{g}_m=\mu_m(\sigma_{\mathrm{x},m})$. Since the reciprocity mismatch coefficients $\bar{g}_m$ of the $m$-th antenna is a nonlinear function of the average transmit power $\sigma_{\mathrm{x},m}$, the calibration coefficients are also related to the power. Therefore, compared with linear reciprocity calibration, the nonlinear reciprocity calibration encounters greater challenges as follows:
\begin{itemize}
	\item \textbf{Challenge 1}: Since the reciprocity mismatch factor varies with the transmit power, it is essential to obtain the relationship $\bar{g}_m=\mu_m(\sigma_{\mathrm{x},m})$ between the nonlinear reciprocity mismatch factor $\bar{g}_m$ and the transmit power $\sigma_{\mathrm{x},m}$.
	\item \textbf{Challenge 2}: From \eqref{eq:gengm}, the function of $g_m$ and $\sigma_{\mathrm{x},m}$ is complex. Hence, it is difficult to determine the expression of the mismatch function $\bar{g}_m=\mu_m(\sigma_{\mathrm{x},m})$.
	\item \textbf{Challenge 3}: As illustrated in Fig. \ref{eq:calibrationcoef}, the relationships between the calibration coefficients are highly nonlinear so that it is difficult to solve the nonlinear calibration coefficients.
\end{itemize}

To overcome these challenges, we propose the multi-power training pilots, the nonlinear reciprocity mismatch polynomial fitting, and the toward optimal nonlinear calibration coefficients for the nonlinear reciprocity calibration.	

\subsection{Pilots Design for Nonlinear Reciprocity Calibration}
From Challenge 1, the reciprocity mismatch is a nonlinear function of the transmit power. To sample the nonlinear function along with the transmit power, we propose multi-power training pilots. Suppose the maximum transmit power of the $m$-th antenna is $\sigma_{\mathrm{max},m}^2$. The power sequences of the calibration signals can be denoted as $\left\lbrace\rho_{\mathrm{c},1},\cdots,\rho_{\mathrm{c},N}\right\rbrace$, where $ \rho_{\mathrm{c},n}=(n\sigma_{\mathrm{max},m}/N)^2 \ (n\in[1:N])$. The calibration signal sequence of $m$-th antenna at the $n$-th power point can be denoted as $\lbrace x_{m}^{(n)}[q]\rbrace_{q\in[1:Q]}$, where $\mathbb{E}\left\lbrace|x_{m}^{(n)}[q]|^2\right\rbrace=\rho_{\mathrm{c},n}$, $ \forall q\in[1:Q]$.

\subsection{Nonlinear Reciprocity Mismatch Factors Estimation}

According to Challenge 2, it is difficult to determine $\mu_m(\sigma_{\mathrm{x},m})$ in practice. Inspired by applying polynomials to characterize the HPA, we employ the polynomials to approximate the function $g_m=t_m\mu(A_{\mathrm{sat},m}/\sigma_{\mathrm{x},m})$ during the reciprocity calibration. Based on \cite{Raich2004Orthogonal}, the function $ \mu_{m}(\sigma_{\mathrm{x},m})={g_m}/{r_m} $ can be expressed by a $\Pi$-order polynomial function as
\begin{equation}
	\mu_{m}(\sigma_{\mathrm{x},m})=\frac{g_m}{r_m}=\sum_{\varpi=0}^{\Pi}\tau_{m,\varpi}\psi_{\varpi}(\sigma_{\mathrm{x},m}),
	\label{eq:polygm}
\end{equation}
where $\tau_{m,\varpi}$ is the polynomial coefficient, and $\psi_{\varpi}(\sigma_{\mathrm{x},m})$ is the orthogonal polynomial given by
\begin{equation}
	\psi_{\varpi}(\sigma_{\mathrm{x},m})=\sum_{l=0}^{\varpi}(-1)^{l+\varpi+2}\frac{(\varpi+l+2)!}{l!(l+1)!(\varpi-l)!}\sigma_{\mathrm{x},m}^l.
\end{equation} Then, we propose an over-the-air training approach based the multi-power pilots to estimate the polynomial coefficient $\tau_{m,\varpi}$.

Let $y_{m,i}^{(n)}$ denote the signal received by the $i$-th antenna. $y_{i,m}$ denotes the received signal by the $m$-th antenna. They are can be given by
\begin{subequations}
	\begin{align}
		{y}_{m,i}^{(n)}[q]=a_0r_ih_{m,i}g_mx_{m}^{(n)}[q]+\tilde{z}_{m,i}^{(n)}[q],\\
		{y}_{i,m}^{(n)}[q]=a_0r_mh_{i,m}g_ix_{i}^{(n)}[q]+\tilde{z}_{i,m}^{(n)}[q],
	\end{align}
\end{subequations}
where $\tilde{z}_{m,i}^{(n)}[q]$ denotes the equivalent noise consisting of the nonlinear distortion and the AWGN at the $i$-th antenna, and $\tilde{z}_{i,m}^{(n)}[q]$ is the equivalent noise at the $m$-th antenna. Then, the polynomial coefficients vector $\boldsymbol{\tau}=[\tau_{1,0},\tau_{1,1},\cdots,\tau_{M,\Pi}]^T$ can be computed by the LS approach as follows.
\begin{prop}[Estimation of the polynomial coefficients]\label{prop:polycoef}
	Assuming that the first polynomial coefficient of the first antenna is a constant, e.g., $\tau_{1,0}=1$, the polynomial coefficients can be computed by
	\begin{equation}
		\hat{\boldsymbol{\tau}}=\left[1,-\bar{\boldsymbol{\psi}}_1^T\mathbf{\Psi}_2^*\left(\mathbf{\Psi}_2^T\mathbf{\Psi}_2^*\right)^{-1}\right]^T,
		\label{eq:solutionalpha}
	\end{equation}
	where $\bar{\boldsymbol{\psi}}_1$ is the first column of $\mathbf{\Psi}$, $\mathbf{\Psi}_2$ consists of the second to the $M\Pi$-th column of $\mathbf{\Psi}$, and $\mathbf{\Psi}$ consists of the received training signals and denoted as \eqref{eq:matrixY}.
\end{prop}
\begin{IEEEproof}
	Based on \eqref{eq:solutionalpha}, since the wireless propagation channel is reciprocal, i.e., $h_{m,i}=h_{i,m}$, the equation can be denoted as $\frac{g_m}{r_m}x_{m}^{(n)}[q]y_{i,m}^{(n)}[q]=\frac{g_i}{r_i}x_{i}^{(n)}[q]y_{m,i}^{(n)}[q]$ by ignoring the noise. By substituting \eqref{eq:polygm} into the equation, it can be further denoted as
	\begin{equation}
		\bar{y}_{m,i}^{(n)}[q]\sum_{\varpi=0}^{\Pi}\tau_{i,\varpi}\psi_{\varpi}(\rho_{\mathrm{c},n})-\bar{y}_{i,m}^{(n)}[q]\sum_{\varpi=0}^{\Pi}\tau_{m,\varpi}\psi_{\varpi}(\rho_{\mathrm{c},n})=0,
		\label{eq:calequation}
	\end{equation}
	where $\bar{y}_{m,i}^{(n)}[q]={y}_{m,i}^{(n)}[q]x_{i}^{(n)}[q]$ and ${y}_{i,m}^{(n)}[q]=\bar{y}_{i,m}^{(n)}[q]x_{m}^{(n)}[q]$. For all $n\in[1:N]$ and $q\in[1:Q]$, \eqref{eq:calequation} holds. Thus, by gathering all the signals and ignoring the noise, the equation can be further denoted as
	\begin{equation}
		\mathbf{\Psi}_{m,i}\boldsymbol{\tau}_{i}-\mathbf{\Psi}_{i,m}\boldsymbol{\tau}_{m}=0,
	\end{equation}
	where $\boldsymbol{\tau}_{i}=[\tau_{i,1},\cdots,\tau_{i,\Pi}]^T$, $\boldsymbol{\tau}_{m}=[\tau_{m,1},\cdots,\tau_{m,\Pi}]^T$, $\mathbf{\Psi}_{m,i}=[\mathbf{\Psi}_{m,i}^{(1)},\cdots,\mathbf{\Psi}_{m,i}^{(N)}]^T$, $\mathbf{\Psi}_{i,m}=[\mathbf{\Psi}_{i,m}^{(1)},\cdots,\mathbf{\Psi}_{i,m}^{(N)}]^T$, $\mathbf{\Psi}_{m,i}^{(n)}=[\bar{y}_{m,i}^{(n)}[1]\boldsymbol{\psi}_{n},\cdots,\bar{y}_{m,i}^{(n)}[Q]\boldsymbol{\psi}_{n}]$,  $\mathbf{\Psi}_{i,m}^{(n)}=[\bar{y}_{i,m}^{(n)}[1]\boldsymbol{\psi}_{n},\cdots,\bar{y}_{i,m}^{(n)}[Q]\boldsymbol{\psi}_{n}]$, and $\boldsymbol{\psi}_{n}=[\psi_{0}(\rho_{\mathrm{c},n}),\cdots,\psi_{\Pi}(\rho_{\mathrm{c},n})]^T$. 
	By stacking the equation of all pairs of antennas into the matrix form, the overall equation can be denoted as
	\begin{equation}
		\mathbf{\Psi}\boldsymbol{\tau}=0,
		\label{eq:pro1}
	\end{equation}
	where $\boldsymbol{\tau}=\left[\boldsymbol{\tau}_{1}^T,\cdots,\boldsymbol{\tau}_{M}^T\right]^T$, and $\mathbf{\Psi}$ is defined as
	\begin{equation}
		\mathbf{\Psi}=\left(
		\begin{matrix}
			\mathbf{\Psi}_{1,2}&-\mathbf{\Psi}_{2,1}&0&\cdots\\
			\mathbf{\Psi}_{1,3}&0&-\mathbf{\Psi}_{3,1}&\cdots\\
			\vdots&\vdots&\vdots&\ddots\\
			0&\mathbf{\Psi}_{2,3}&-\mathbf{\Psi}_{3,2}&\cdots\\
			\vdots&\vdots&\vdots&\ddots
		\end{matrix}
		\right),
		\label{eq:matrixY}
	\end{equation}
	To exclude the trivial all-zero solution to \eqref{eq:pro1}, $\tau_{1,0}$ is assumed to be known previously and set to $1$. Then, \eqref{eq:pro1} can be further written as
	\begin{equation}
		\mathbf{\Psi}_2\boldsymbol{\tau}_{\mathrm{c}}+\bar{\boldsymbol{\psi}}_1=0,
		\label{eq:lspro}
	\end{equation} 
	where $\boldsymbol{\tau}_{\mathrm{c}}$ consists of the second to the $\Pi$-th row of $\boldsymbol{\tau}$, $\bar{\boldsymbol{\psi}}_1$ is the first column of $\mathbf{\Psi}$ and $\mathbf{\Psi}_2$ consists of the second to the $M\Pi$-th column of $\mathbf{\Psi}$.
	\eqref{eq:lspro} can be solved by the least square algorithm and its solution can be given by
	\begin{equation}
		\hat{\boldsymbol{\tau}}_{\mathrm{c}}=-(\mathbf{\Psi}_2^H\mathbf{\Psi}_2)^{-1}\mathbf{\Psi}_2^H\bar{\boldsymbol{\psi}}_1.
	\end{equation}
	As a result, the coefficient vector $\boldsymbol{\tau}=[\tau_1,\hat{\boldsymbol{\tau}}_{\mathrm{c}}^T]^T$ can be denoted as \eqref{eq:solutionalpha}.
\end{IEEEproof}

\subsection{Computing the Reciprocity Calibration Coefficients}\label{sec:cal}
In terms of \eqref{eq:caequprimalnl}, only the amplitude of the calibration coefficients changes the output of HPA. Hence, the amplitude and phase of the calibration coefficient can be computed independently. By substituting \eqref{eq:polygm} into \eqref{eq:caequprimalnl}, the amplitude and phase should, respectively, satisfy
\begin{align}
	&|c_m|\bar{\mu}_m(|c_m|)=|c_m|\bar{\mu}_i(|c_i|),\label{eq:amcal}\\
	&\angle c_m+\angle \mu_m(|c_m|)=\angle c_i+\angle \mu_i(|c_i|),\label{eq:phcal}
\end{align}
where $\bar{\mu}_m(|c_m|)$ is the amplitude of $\mu_m(|c_m|\sigma_{\mathrm{x},m})$ and can be denoted as
\begin{equation}
	\begin{split}
		\bar{\mu}_m(|c_m|)=\sqrt{\mu_{\mathrm{r},m}^2(|c_m|\sigma_{\mathrm{x},m})+\mu_{\mathrm{i},m}^2(|c_m|\sigma_{\mathrm{x},m})}.
	\end{split}
	\label{eq:amfunction}
\end{equation}
In \eqref{eq:amfunction}, $\mu_{\mathrm{r},m}(\sigma_{\mathrm{x},m})=\sum_{\varpi=0}^{\Pi}\hat{\tau}_{\mathrm{r},m,\varpi}\psi_{\varpi}(\sigma_{\mathrm{x},m})$,   $\mu_{\mathrm{i},m}(\sigma_{\mathrm{x},m})=\sum_{\varpi=0}^{\Pi}\hat{\tau}_{\mathrm{i},m,\varpi}\psi_{\varpi}(\sigma_{\mathrm{x},m})$, where $\hat{\tau}_{\mathrm{r},m,\varpi}$ is the real part of $\hat{\tau}_{m,\varpi}$, and $\hat{\tau}_{\mathrm{i},m,\varpi}$ denotes the imaginary part of $\hat{\tau}_{m,\varpi}$.
Therefore, the amplitude and phase of calibration coefficients can be computed by solving \eqref{eq:amcal} and \eqref{eq:phcal}, respectively.

As mentioned by Challenge 3, it is difficult to solve the calibration coefficients from \eqref{eq:amcal}. We can find that there are infinitely many solutions to \eqref{eq:amcal}, even including an all-zero solution which can make the system fail. Due to the complex expression of the function $\mu_{m}(x)$, it is difficult to find the calibration coefficients exactly satisfying the total power constraint and the maximum power constraint based on \eqref{eq:amcal}. To address the issues, an optimization problem is formulated to solve the equations efficiently and find the optimal calibration coefficients. The problem seeks to maximize the downlink achievable rate with maximum transmit power constraints, i.e., $\sum_{m=1}^{M}|c_m|^2\sigma_{\mathrm{x},m}^2\leq\rho_{\mathrm{t}}$ and $|c_m|\sigma_{\mathrm{x},m}\leq \sigma_{\mathrm{max},m}$, and reciprocity constraints in \eqref{eq:amcal}. Thus, the problem can be formulated as
\begin{equation}
	\begin{split}
		\mathcal{P}_1:~&\max_{|c_m|}\ \gamma_k\\
		\mathrm{s.t.}\ &\mathcal{C}_1:\sum_{m=1}^{M}|c_m|^2\sigma_{\mathrm{x},m}^2\leq\rho_{\mathrm{t}},\\
		&\mathcal{C}_2:\varphi_m(|c_m|)=\varphi_i(|c_i|),\\
		&\mathcal{C}_3:0\leq |c_m|\leq c_{\mathrm{max},m},
	\end{split}
\end{equation}
where $\varphi_m(|c_m|)=|c_m|\bar{\mu}_m(|c_m|)$ and $c_{\mathrm{max},m}=\sigma_{\mathrm{max},m}/\sigma_{\mathrm{x},m}$.

Since both the objective and $\mathcal{C}_2$ are non-convex, $\mathcal{P}_1$ is non-convex and difficult to solve. To solve $\mathcal{P}_1$ efficiently, we reformulate $\mathcal{P}_1$ as an equivalent convex problem as follows.
\begin{prop}[Equivalent convex optimization]\label{prop:optimalconvex}
	By substituting \eqref{eq:caequprimal} into the objective and relaxing the equation constraint, $\mathcal{P}_1$ can transformed into an equivalent convex problem as
	\begin{equation}
		\begin{split}
			\mathcal{P}_2:~&\max_{|c_m|,g_{\mathrm{0}}}\ \ g_{\mathrm{0}}\\
			\mathrm{s.t.}\ \ &\mathcal{C}_1, \mathcal{C}_3,\\
			&\mathcal{C}_4:0\leq g_0\leq\varphi_{m}(|c_m|),\ \forall m\in [1:M].
		\end{split}
		\label{eq:convexpro}
	\end{equation}
\end{prop}
\begin{IEEEproof}
	From the theoretical analysis, the distortion is negligible in practical systems and can be regarded as the equivalent noise. By substituting \eqref{eq:caequprimal} into the SINDR in \eqref{eq:sinrzf} and focusing on the nonlinear reciprocity mismatch at the BS side, the SINDR can be further denoted as $\gamma_{k}=(M-K)\rho_{\mathrm{t}}a_0g_0^2|\mathrm{tr}\left\lbrace \mathbf{R}^*\right\rbrace|^2/(\mathrm{tr}\left\lbrace \mathbf{\Phi}^{-2}\right\rbrace M\mathrm{tr}\left\lbrace \mathbf{RR}^*\right\rbrace\sigma_{\mathrm{n}}^2)$, where $g_0=\varphi_{m}(|c_m|)$. Since the quadratic function is monotonically increasing when $g_0\geq 0$, the objective of $\mathcal{P}_1$ can be replaced by $g_0$ and the constraint $\mathcal{C}_2$ can be rewritten as $g_0=\varphi_{m}(|c_m|),\ \forall m\in[1:M]$. 
	
	Since $\varphi_{m}(|c_m|)$ is a nonlinear function of $|c_m|$, the equation constraint $\mathcal{C}_2$ is nonconvex. Thanks to the concavity of $\varphi_{m}(|c_m|)$, the constraint $\mathcal{C}_2$ becomes convex by relaxing the equation, i.e., $g_0\leq\varphi_{m}(|c_m|)$. Hence, $\mathcal{P}_1$ can be reformulated as $\mathcal{P}_2$ as \eqref{eq:convexpro}. Because both the objective function and the function $g0-\varphi_{m}(|c_m|)$ are monotonically increasing with $g_0$, the problem $\mathcal{P}_2$ after relaxation is equivalent to the primal problem $\mathcal{P}_1$ \cite{boyd2004convex}. In other words, $\mathcal{P}_2$ is equivalent to $\mathcal{P}_1$.
\end{IEEEproof}

As $\mathcal{P}_2$ is a convex optimization problem, it has a unique maximum. The optimal solution can be obtained by some math tools, e.g., the interior-point method \cite{boyd2004convex}, but the performance of such method decreases rapidly as the antenna number $M$ increases. Inspired by the sequential linear programming (SLP) \cite{Powell1978A}, we propose an efficient algorithm, which can achieve the same performance of the interior-point method.

By employing the Taylor expansion centered at $|c_m^{l-1}|$, the nonlinear function $\varphi_{m}(|c_m|)$ can be approximated as
\begin{equation}
	\begin{split}
		\varphi_{m}(|c_m|)\approx \tilde{\varphi}_m(|c_m^{l-1}|,|c_m|)
		= \varphi_{m}(|c_m^{l-1}|)+\varphi_{m}'(|c_m^{l-1}|)(|c_m|-|c_{m}^{l-1}|),
	\end{split}
	\label{eq:taylorexp}
\end{equation}
where $\left\lbrace |c_{m}^{l-1}|\right\rbrace_{m\in[1:M]}$ is the solution to the $(l-1)$-th iteration optimization. Then, the subproblem of the $l$-th iteration can be denoted as
\begin{equation}
	\begin{split}
		\mathcal{P}_3:~&\max_{f_m,g_{\mathrm{0}}}\ \ g_{\mathrm{0}}\\
		\mathrm{s.t.}\ \ &\mathcal{C}_1, \mathcal{C}_2,\\
		&\mathcal{C}_5:0\leq g_0\leq \tilde{\varphi}_m(|c_m^{l-1}|,|c_m|),\ \forall m\in [1:M].
	\end{split}
\end{equation}
$\mathcal{P}_3$ is a linear programming and the solution can be given as follows.
\begin{prop}[Solution to the $l$-th subproblem]
	The problem $\mathcal{P}_3$ is convex and has a unique solution as
	\begin{equation}
		\begin{cases}
			\bar{c}_m^{l}=\frac{g_0^l-\chi_m^{l-1}}{\varphi_m'(|c_m^{l-1}|)},\\
			g_0^{l}=\min\left\lbrace\tilde{\varphi}_m(|c_m^{l-1}|,c_{\mathrm{max},m}),\hat{g}_0\right\rbrace,
		\end{cases}
		\label{eq:optsol}
	\end{equation}
	where $\hat{g}_0$ is the solution to the quadratic equation $\sum_{m=1}^{M}\left[\frac{(g_0-\chi_m^{l-1})\sigma_{\mathrm{x},m}}{\varphi_m'(|c_m^{l-1}|)}\right]^2=\rho_{\mathrm{t}}$ and can be solved by the quadratic formula, and $\chi_m^{l-1}=\varphi_m(|c_m^{l-1}|)-\bar{\varphi}_m'(|c_m^{l-1}|)|c_m^{l-1}|$.
\end{prop}
\begin{IEEEproof}
	The solution to $\mathcal{P}_3$ can be solved by using the Lagrange method and Karush-Kuhn-Tucker conditions \cite{boyd2004convex}.
\end{IEEEproof}

In \eqref{eq:optsol}, to ensure that $\bar{c}_m^l$ is non-negative, $g_0^l$ should be larger than $\max_{m}\left\lbrace\chi_m^{l-1}\right\rbrace$, i.e.,
\begin{equation}
	\begin{split}
		&\min_{m}\left\lbrace\tilde{\varphi}_m(|c_m^{l-1}|,c_{\mathrm{max},m})\right\rbrace> \max_{m}\left\lbrace\chi_m^{l-1}\right\rbrace,\\
		&\sum_{m=1}^{M}\left[\frac{\max_{m}\left\lbrace\chi_m^{l-1}\right\rbrace-\chi_m^{l-1}}{\varphi_m'(|c_m^{l-1}|)\sigma_{\mathrm{x},m}^{-1}}\right]^2<\rho_{\mathrm{t}}.
	\end{split}
	\label{eq:inequal}
\end{equation}
Therefore, we propose a simple line search approach to choose a step size $\varsigma\in(0,1]$ ensuring that the inequalities in \eqref{eq:inequal} are true and $\min_{m}\left\lbrace \varphi_m(|c_m^{l}|)\right\rbrace$ is non-decreasing. Assume that the $\lbrace |c_m^{l-1}|\rbrace_{m\in[1:M]}$ in the $(l-1)$-th iteration can guarantee the inequalities in \eqref{eq:inequal}. The step size $\alpha$ can be the largest element in $\left\lbrace\varrho^j\right\rbrace_{j=0,1,\cdots}$ satisfying $\min_{m}\left\lbrace \varphi_m(|c_m^{l-1}|)\right\rbrace\leq\min_{m}\left\lbrace \varphi_m(|c_m^{l}|)\right\rbrace$ and the inequalities in \eqref{eq:inequal}, where $\varrho\in(0,1)$, and $|c_m^{l}|=(1-\varsigma)|c_m^{l-1}|+\varsigma\bar{c}_m^{l}$. Finally, the SLP approach for computing the nonlinear calibration coefficients is summarized as Algorithm \ref{alg:SLPforcal}.

\begin{algorithm}
	\caption{SLP for computing the calibration coefficients.}
	\label{alg:SLPforcal} 
	\begin{algorithmic}[1]
		\REQUIRE
		$|c_m^{0}|=0\ \forall m\in[1:M]$, $\epsilon$.
		\ENSURE
		The nonlinear reciprocity calibration coefficients $\left\lbrace\bar{c}_m\right\rbrace_{m\in[1:M]}$.
		\REPEAT
		\STATE Compute $\varphi_m(|c_m^{l-1}|)$, $\varphi_m'(|c_m^{l-1}|)$, and $\chi_m^{l-1}$, $\forall m\in[1:M]$.
		\STATE Solve the quadratic equation $\sum_{m=1}^{M}\left[\frac{(g_0-\chi_m^{l-1})\sigma_{\mathrm{x},m}}{\varphi_m'(|c_m^{l-1}|)}\right]^2=\rho_{\mathrm{t}}$ by the quadratic formula.
		\STATE Set $ g_0^l=\min\left\lbrace\tilde{\varphi}_m(|c_m^{l-1}|,c_{\mathrm{max},m}),\hat{g}_0\right\rbrace $.
		\STATE Set $\bar{c}_m^l=\frac{g_0^{l}-\chi_m^{l-1}}{\varphi_m'(|c_m^{l-1}|)}$, $\forall m\in[1:M]$.
		\STATE Step size rule: Set $\Delta c_m^{l}=\bar{c}_m^l-|c_m^{l-1}|$ and choose a $\varrho\in(0,1)$. Let $\varsigma^{(l)}$ be the largest element in $\left\lbrace\varrho^j\right\rbrace_{j=0,1,\cdots}$ satisfying $\min_{m}\left\lbrace \varphi_m(|c_m^{l-1}|)\right\rbrace\leq\min_{m}\left\lbrace \varphi_m(|c_m^{l-1}|+\varsigma^{(l)}\Delta c_{m}^l)\right\rbrace$ and the inequalities denoted in \eqref{eq:inequal}.
		\STATE Set $|c_m^{l}|=|c_m^{l-1}|+\varsigma^{(l)}\Delta c_m^l$, $\forall m\in[1:M]$.
		\UNTIL $\frac{1}{M}\sum_{m=1}^{M}\left||c_m^{l}|-|c_m^{l-1}|\right|<\epsilon$.
	\end{algorithmic}
\end{algorithm}

\begin{rem}[Convergence analysis]
	$\mathcal{P}_2$ equivalently maximizes $\min_{m}\left\lbrace \varphi_m(|c_m|)\right\rbrace$. Since $\min_{m}\left\lbrace \varphi_m(|c_m|)\right\rbrace$ is non-decreasing during the iterations and is a bounded function, $(\bar{c}_1,\cdots,\bar{c}_M)$ is a limit point of the iterations in Algorithm \ref{alg:SLPforcal}, which yields $\lim_{l\rightarrow \infty}\min_{m}\left\lbrace \varphi_m(|c_m^l|)\right\rbrace=\min_{m}\left\lbrace \varphi_m(\bar{c}_m)\right\rbrace$. According to \eqref{eq:taylorexp}, the limit point $(\bar{c}_1,\cdots,\bar{c}_M)$ is likewise the stationary point of $\mathcal{P}_2$, i.e., $\lim_{l\rightarrow \infty}\min_{m}\left\lbrace\varphi(|c_m|^{l})\right\rbrace= \min_{m}\left\lbrace\varphi(|c_m^*|)\right\rbrace$, where $\left\lbrace |c_m^*|\right\rbrace_{m\in[1:M]}$ is the optimal solution to $\mathcal{P}_2$.
\end{rem}

Further, the phases of calibration coefficients can be computed by solving the equations denoted in \eqref{eq:phcal}. Since the variables are more than the equations, the equations are underdetermined and have infinitely many solutions. Since the phases of calibration coefficients unrelated to the average transmit power, any particular solution to the equations can make the system work well. Therefore, we can give a particular solution denoted as
\begin{equation}
	\begin{split}
		\angle c_m&=-\angle \mu_n(\sigma_{\mathrm{x},m})=-\arctan\left(\frac{\mu_{\mathrm{r},m}(|c_m|\sigma_{\mathrm{x},m})}{\mu_{\mathrm{i},m}(|c_m|\sigma_{\mathrm{x},m})}\right),\ \forall m\in[1:M].
	\end{split}
	\label{eq:phasecoef}
\end{equation}
Thus, the precoding matrix of ZF for implementing the nonlinear reciprocity calibration can be denoted as $\mathbf{W}_{c}=\mathrm{diag}(\mathbf{c})\mathbf{W}$.

\subsection{Calibration Process, Overhead and Complexity Analysis}

\begin{algorithm}
	\caption{Nonlinear reciprocity calibration}
	\label{alg:nonlinearrc}
	\begin{itemize}
		\item \textbf{Step 1 (Transmit training pilots):} The antenna $m$ transmits the training pilots $\left\lbrace x_{m}^{(n)}[1],\cdots,x_{m}^{(n)}[Q]\right\rbrace$, and antenna $i\ (i\neq m)$ receives the training signals $y_{m,i}^{(n)}[q]$, $\forall m,i\in [1:M],\ n\in[1:N],\ q\in[1:Q]$.
		\item \textbf{Step 2 (Estimate the polynomial coefficients):} The baseband processor gathers all the received training signals and formulates the matrix $\mathbf{\Psi}$ denoted in \eqref{eq:matrixY}. Then, the polynomial coefficients can be estimated by \eqref{eq:solutionalpha}.
		\item \textbf{Step 3 (Compute the nonlinear calibration coefficients):} The baseband processor formulates the amplitude function $\bar{\mu}(\sigma_{\mathrm{x},m})$ denoted in \eqref{eq:amfunction}. Then, the amplitude of the calibration coefficients can be computed by solving the optimization problem $\mathcal{P}_2$ with Algorithm \ref{alg:SLPforcal}. The phase of the calibration coefficients can be computed by \eqref{eq:phasecoef}.
	\end{itemize}
\end{algorithm}

The overall process of the nonlinear reciprocity calibration can be described as Algorithm \ref{alg:nonlinearrc}. According to Algorithm \ref{alg:nonlinearrc}, it can be seen that the overhead is caused by transmitting the training signals, and the computational complexity mainly results from computing the polynomial coefficients and the nonlinear calibration coefficients, respectively. The training overhead is caused by learning the system coefficients and can be defined as the number of transmitting the training pilots. As for the computational complexity, we focus on the complexity resulting from the multiplication. Then, the training overhead and the computational complexity for the polynomial nonlinear reciprocity calibration are given as follows.

\begin{rem}[Overhead and Computational complexity]
	For the polynomial nonlinear reciprocity calibration, the total training overhead is $ MNQ $. The computational complexity of computing the polynomial coefficients can be given
	by $ \mathcal{O}(M^3\Pi^3+M^2Q^2N^2\Pi) $ and the computational complexity of computing calibration coefficients by Algorithm \ref{alg:SLPforcal} is $ \mathcal{O}(M\Pi) $. According to \cite{boyd2004convex}, the complexity of the interior-point method for solving $\mathcal{P}_2$ can be given by $\mathcal{O}(\Pi M^{1.5}\log(M)+M^{3.5}\log M)$. Hence, with the same solution accuracy, the proposed Algorithm \ref{alg:SLPforcal} is less complex than the interior-point method.
	%
	%
\end{rem}

\section{Simulation Results and Discussions}
In this section, we provide simulation results for the multi-user massive MIMO system in the presence of the nonlinear reciprocity mismatch to verify the analytical results and to show the performance of the proposed calibration approach. For easy presentation, RC denotes reciprocity calibration, and NRC represents nonlinear reciprocity calibration.

The system parameters for simulations are set as follows. The cell radius is normalized to $ 1 $ and the minimum distance between BS and UE is set to $ 0.01 $ \cite{Nie2019A}. The BS is equipped with $M=256$ antennas and is deployed at the center of the cell. There are $K=20$ single-antenna UEs served by the BS simultaneously. The large-scale path loss between the $k$-th UE and BS is modeled as $\phi_k=\zeta d_{\mathrm{h},k}^{-\xi}$, where $\zeta$ is the path gain at the reference distance of the far-field area, $d_{\mathrm{h},k}$ is the distance between the BS and UE $k$, and $\xi$ is the path loss exponent \cite{Nie2020Nie2020Relaying}. In the simulation, $\zeta$ is set to $-20$ dB and $\xi$ equals to $3.7$. The variance of AWGN is set to $\sigma_{\mathrm{n}}^2=1$, and $\rho_{\mathrm{t}}$ denotes the average transmit SNR. The small amplification gain $a_0$ is equal to $10$ dB. Further, both the IBO and the mismatch coefficients are differently set in each simulation.


\subsection{Impact of Nonlinear Reciprocity Mismatch} 

\begin{figure}
	\centering
	\begin{minipage}{0.49\linewidth}
		\includegraphics[width=\linewidth]{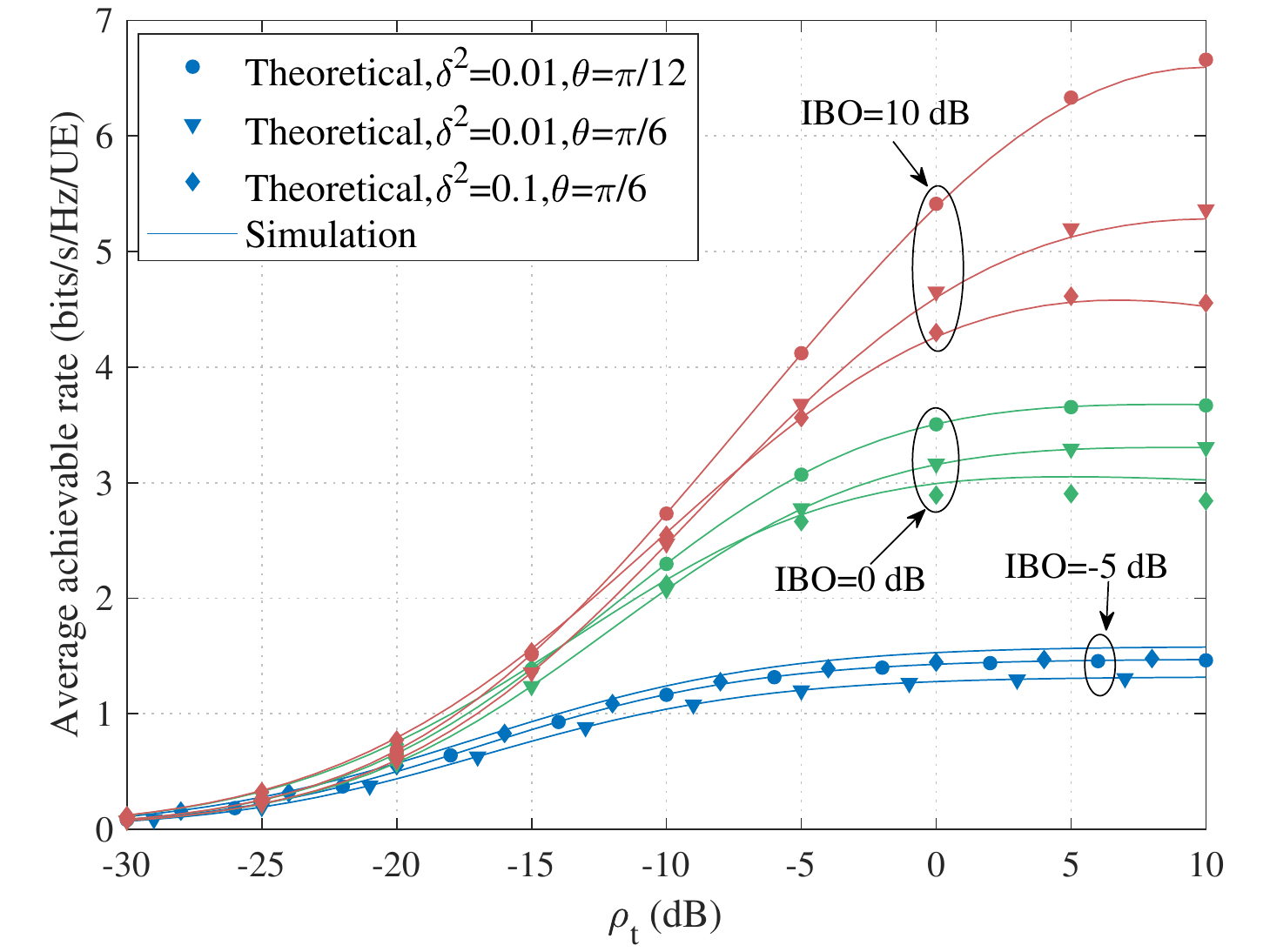}
		\caption{Average achievable rate versus average transmit power $\rho_{\mathrm{t}}$.}
		\label{fig:rateVsRhot}		
	\end{minipage}
	\hfill
	\begin{minipage}{0.49\linewidth}
		\centering
		\includegraphics[width=\linewidth]{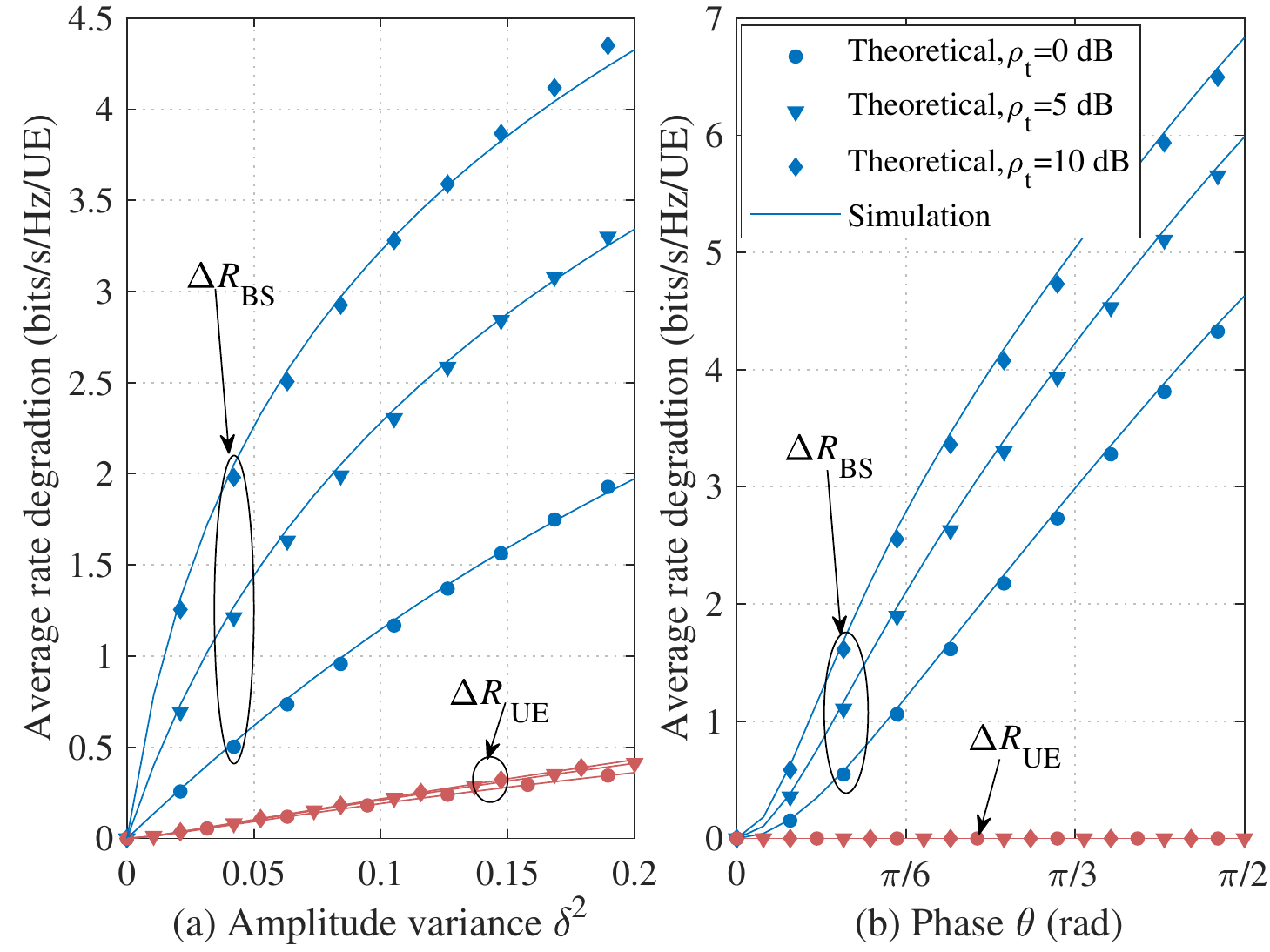}
		\caption{Average achievable rate versus reciprocity mismatch coefficients $\delta^2$ and $\theta$.}
		\label{fig:RateVSAmPhi}		
	\end{minipage}
\end{figure}

Fig. \ref{fig:rateVsRhot} illustrates the downlink average achievable rate as a function of the average transmit SNR $\rho_{\mathrm{t}}$ for different values of IBO, $\delta^2$, and $\theta$. The amplitudes of $a_m$, $t_m$, $r_m$, $u_k$, and $v_k$ are distributed as $\left\lbrace \ln a_m,\ln|t_m|,\ln|r_m|,\ln|u_k|,\ln|v_k|\right\rbrace\sim\mathcal{N}(0,\delta^2)$, and phases of $t_m$, $r_m$, $u_k$, and $v_k$ are distributed as $\left\lbrace \angle t_m,\angle r_m,\angle u_k,\angle v_k\right\rbrace\sim\mathcal{U}(-\theta,\theta)$. From the figure, the theoretical SINDR and achievable rate denoted in Proposition \ref{theo:ratezf} and Proposition \ref{coro:arate} is accurate. The achievable rate increases with the transmit SNR and approaches an upper limit when the SNR is large. Further, as the IBO increases or the mismatch parameters decrease, the upper limit of the achievable rate increases. This implies that the rate limitation is caused by both the limitation of the HPA and the reciprocity mismatch, which is consistent with analytical results.

The impacts of the amplitude and phase reciprocity mismatch on the average achievable rate degradation are shown in Fig. \ref{fig:RateVSAmPhi}a and Fig. \ref{fig:RateVSAmPhi}b, respectively. In the two figures, $\Delta R_{\mathrm{BS}}$ (see \eqref{eq:ratebs}) denotes the achievable rate degradation only caused by the nonlinear reciprocity mismatch at the BS side, while $\Delta R_{\mathrm{UE}}$ (see \eqref{eq:ratelossUE}) represents the achievable rate degradation only resulting from the mismatch at the UE side. From Fig. \ref{fig:RateVSAmPhi}a, both $\Delta R_{\mathrm{BS}}$ and $\Delta R_{\mathrm{UE}}$ increases with $\delta^2$ increasing. In Fig. \ref{fig:RateVSAmPhi}b, only $\Delta R_{\mathrm{BS}}$ increases when $\theta$ increases, and $\Delta R_{\mathrm{UE}}=0$, which implies the phase mismatch at the UE side does not degrade the achievable rate. Further, it also can be seen that only $\Delta R_{\mathrm{BS}}$ increases with the transmit SNR $\rho_{\mathrm{t}}$. The nonlinear reciprocity mismatch at the BS side causes much more severe performance loss than the nonlinear reciprocity mismatch at the UE side. Therefore, the reciprocity calibration at the BS side is very essential for the TDD multi-user massive MIMO system.

\begin{figure}
	\centering
	\begin{minipage}{0.49\linewidth}
		\includegraphics[width=\linewidth]{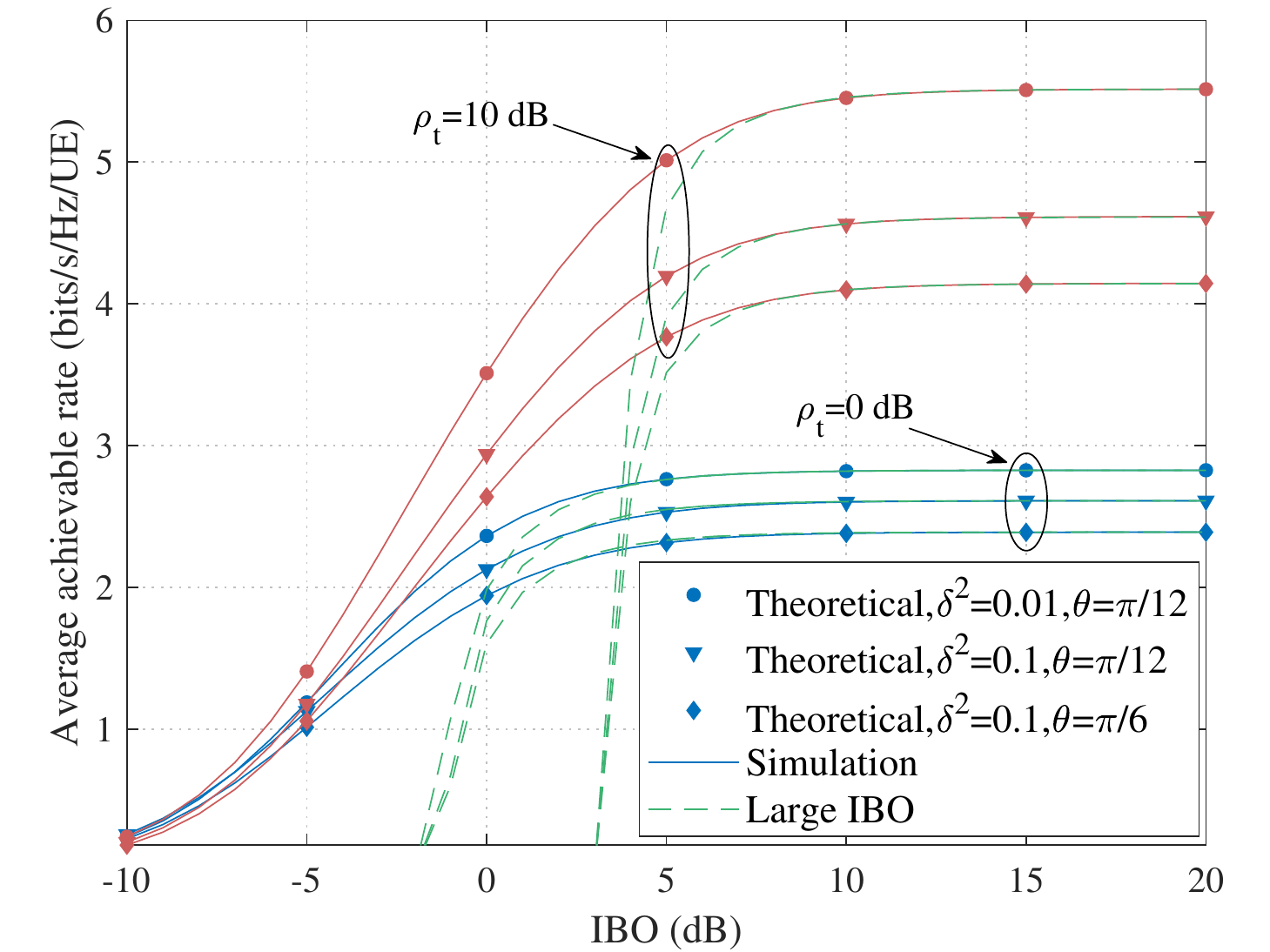}
		\caption{Average achievable rate versus IBO with different mismatch parameters.}
		\label{fig:rateVSIBO}
	\end{minipage}
	\hfill
	\begin{minipage}{0.49\linewidth}
		\includegraphics[width=\linewidth]{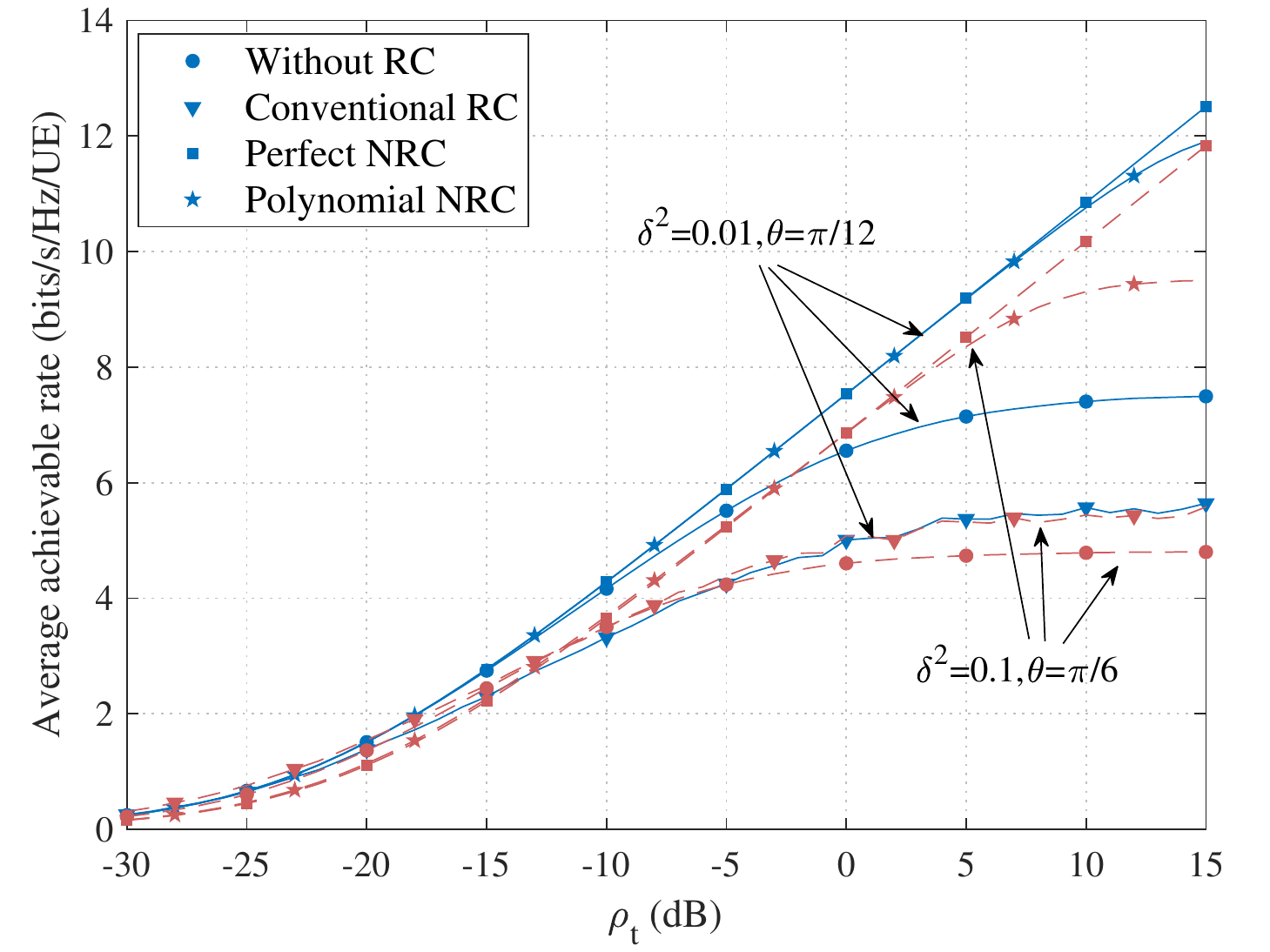}
		\caption{Average achievable rate after reciprocity calibration versus average transmit power $\rho_{\mathrm{t}}$.}
		\label{fig:RCRateVsRhot}
	\end{minipage}
\end{figure}

Further, the average achievable rate versus the IBO of the BS is demonstrated in Fig. \ref{fig:rateVSIBO} with the transmit SNR $\rho_{\mathrm{t}} $ set to $0$ dB and $10$ dB. As seen from the figure, the achievable rate increases when IBO increases. In the small IBO region, the achievable rates are almost the same for the different system coefficients. This indicates that the poor amplification ability of the HPA greatly limits system performance. In the large IBO region, the achievable rate approaches an upper limit. As the transmit SNR $\rho_{\mathrm{t}}$ increases or the mismatch parameters decrease, the upper limit of the achievable rate increases, which implies that the transmit power and the reciprocity mismatch limit the system performance. Further, the impact of the reciprocity mismatch on the achievable rate becomes greater when both the transmit SNR and IBO are large. These results indicate that the nonlinearity of HPAs intensifies the reciprocity mismatch, which is consistent with the theoretical results in \eqref{eq:highpowerZF}.

\subsection{Performance of Nonlinear Reciprocity Calibration}

Fig. \ref{fig:RCRateVsRhot} demonstrates the average achievable rate versus the transmit SNR $\rho_{\mathrm{t}}$ with different calibration approaches. Note that the perfect NRC is the performance benchmark of the nonlinear reciprocity calibration. In the simulation, the polynomial order is set to $\Pi=5$ and the IBO is $10$ dB. The training pilot length $N$ and $Q$ are set to $5$ and $10$, respectively. From the figure, it can be seen that the achievable rate increases with the transmit SNR. In the small SNR regime, the system with any reciprocity calibration approach almost has the same performance as the system without the reciprocity calibration. In the high SNR regime, the achievable rate with the nonlinear reciprocity calibration is much larger than both the rate without calibration and the rate with conventional calibration. The results imply that the reciprocity calibration is more essential to the TDD system working at the high SNR regime. The performance of the polynomial nonlinear reciprocity calibration is very closed to the perfect nonlinear reciprocity calibration. When the transmit SNR is large, the performance of the polynomial NRC is less than the perfect NRC. This is because $|\hat{\mu}_{m}(\sigma_{\mathrm{x},m})-{\mu}_{m}(\sigma_{\mathrm{x},m})|\sigma_{\mathrm{x},m}$ becomes larger when the transmit power increases, where $\hat{\mu}_{m}(\sigma_{\mathrm{x},m})$ is the learned nonlinear mismatch function from the training, and ${\mu}_{m}(\sigma_{\mathrm{x},m})$ denotes the actual nonlinear mismatch function.

\begin{figure}
	\centering
	\begin{minipage}{0.49\linewidth}
		\includegraphics[width=\linewidth]{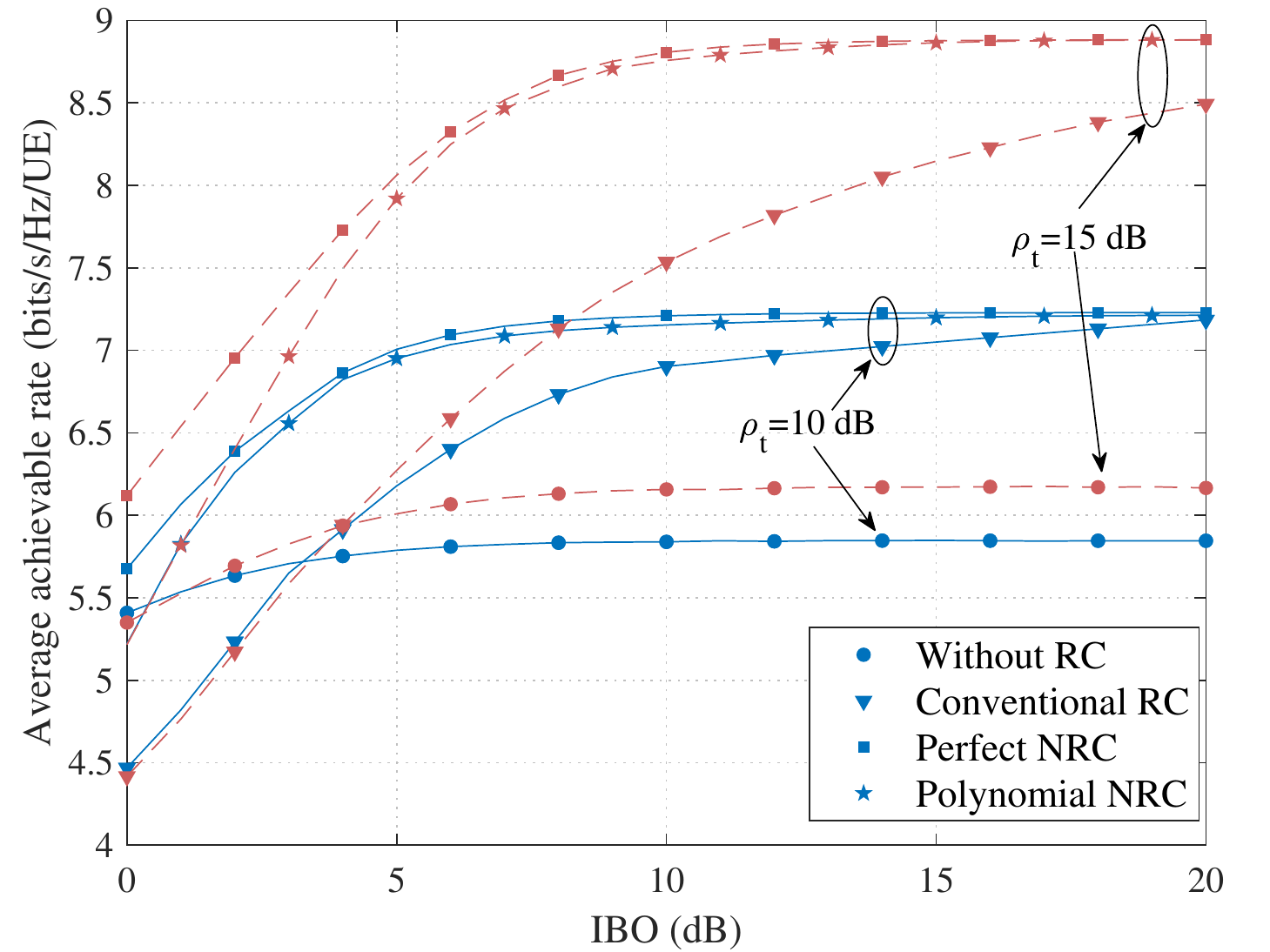}
		\caption{Average achievable rate after reciprocity calibration versus IBO.}
		\label{fig:RCRateVSIBO}
	\end{minipage}
	\hfill
	\begin{minipage}{0.49\linewidth}
		\includegraphics[width=\linewidth]{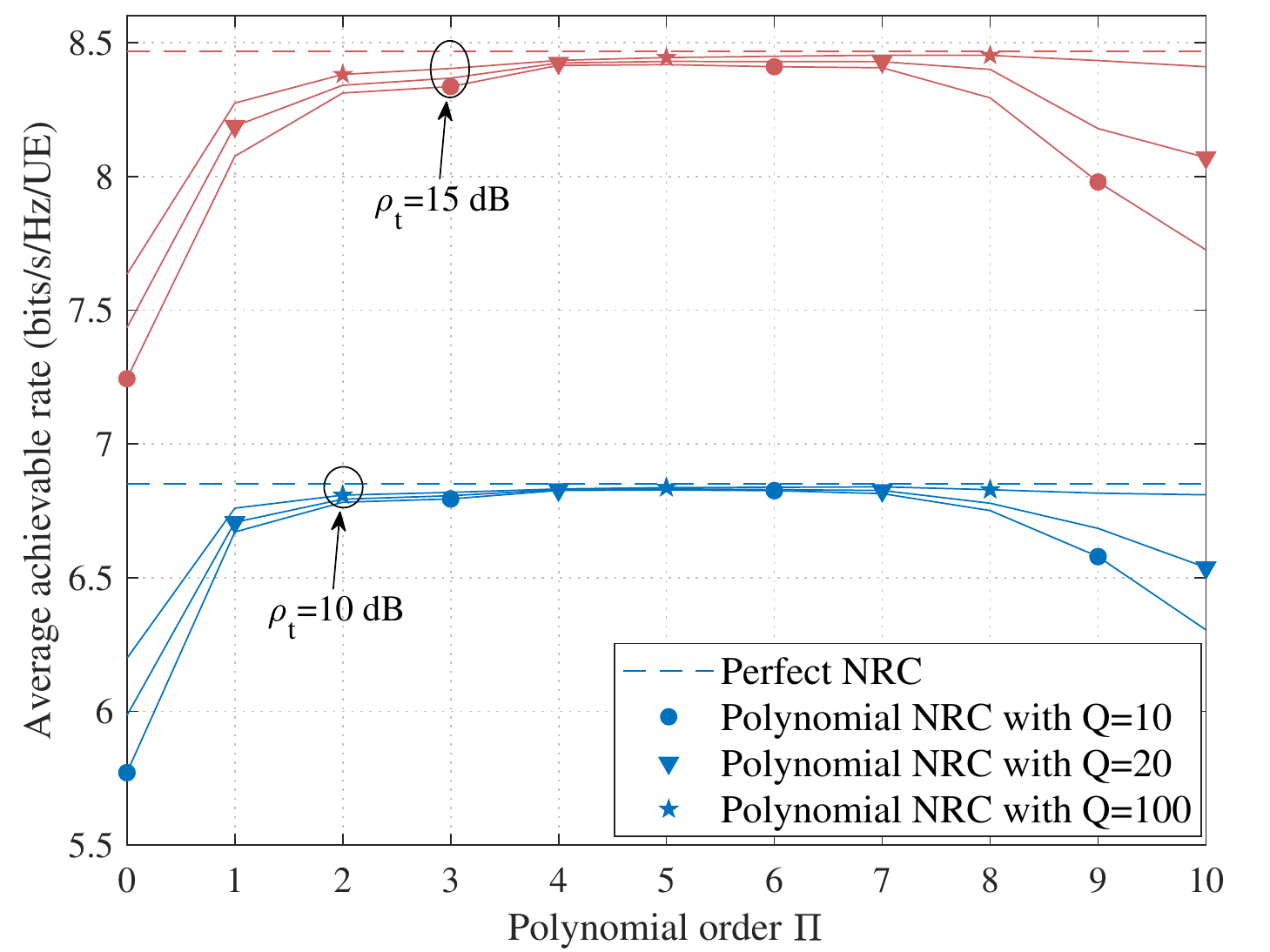}
		\caption{Average achievable rate after reciprocity calibration versus the polynomial order $\Pi$.}
		\label{fig:RCRateVSOrder}
	\end{minipage}
\end{figure}

The average achievable rate after the reciprocity calibration versus the IBO is illustrated in Fig. \ref{fig:RCRateVSIBO} with $\rho_{\mathrm{t}}$ set to $10$ dB and $15$ dB. The mismatch coefficients $\left\lbrace\delta^2,\theta\right\rbrace$ are set to $\left\lbrace 0.05,\pi/6\right\rbrace$. From the figure, the average achievable rate after the reciprocity calibration increases with the IBO. In the small IBO regime, the achievable rate of the polynomial NRC is smaller than the perfect NRC, and it gradually approaches the rate of the perfect NRC. At the large IBO regime, the polynomial NRC has the same performance as the perfect NRC. This is because the estimation error of the polynomial coefficients decreases with the increase of IBO. Further, the conventional calibration performs poorly when IBO is small, and it performs better as the IBO increases. At the very large IBO regime, the performance of the conventional calibration approaches the NRC. This is because the nonlinearity of HPA disappears when IBO is very large, and the TDD system suffers the linear reciprocity mismatch.

Finally, the relationship between the achievable rate and the polynomial order is shown in Fig. \ref{fig:RCRateVSOrder}. The IBO is set to $5$ dB, and the mismatch parameters $\left\lbrace \delta^2,\theta\right\rbrace$ are set to $\left\lbrace 0.1,\pi/6\right\rbrace$. The order $\Pi=0$ denotes the conventional RC approach. The figure shows that when Q is small, as the polynomial order increases, the achievable rate first increases and then decreases. This is because the received training signal is noisy when $Q$ is small. The high-order polynomial fitting by using the noisy data suffers over-fitting, which makes the learned polynomial function perfectly fit the noisy training data, but fail to fit the actual function. When the pilot length $Q$ is large, the performance of the polynomial NRC always approaches the perfect NRC as the polynomial order is large. This is because the impact of the noise on the polynomial fitting gradually decreases and even vanishes, as the length of the training pilots increases. Consequently, to reduce the computational complexity and achieve a good performance of the polynomial nonlinear reciprocity calibration, the polynomial order should not be very large.

\section{Conclusions}
In this paper, we have studied the nonlinear reciprocity mismatch of the multi-user massive MIMO system, including the impact analysis and the nonlinear reciprocity calibration. By modeling the transmit RF gain as a nonlinear function of the transmit power, we derived the closed-form expression of the ergodic achievable rate with the nonlinear reciprocity mismatch. Based on the closed-form achievable rate, the performance loss caused by the nonlinear mismatch at the BS side and the UE side was presented, respectively. The analytical results revealed that the impact of the mismatch at the BS side was much severer than that at the UE side. To further analyze the impact of the nonlinear mismatch at the BS side, we considered a special case where the IBO was larger than zero, which demonstrated that the nonlinearity exacerbated the reciprocity mismatch. Then, we proposed a novel nonlinear reciprocity calibration approach for the BS. During the calibration, the nonlinear relationship between the mismatch factor and the transmit power was approximated by the polynomial fitting, and the polynomial coefficients were estimated by the over-the-air training. Finally, to compute the nonlinear calibration coefficients efficiently, we formulated an auxiliary optimization problem and proposed a fast algorithm to solve it. Due to the low complexity of the algorithm, the nonlinear NRC was easy to implement in the actual system.

\appendices

\section{Proof of Proposition \ref{theo:ratezf}}\label{append:rateZF}
According to \eqref{eq:defbetazf}, the closed-form expression of the normalization scalar $\beta_{\mathrm{ZF}}$ of the ZF precoding scheme can be given by
\begin{equation}
	\begin{split}
		\beta_{\mathrm{ZF}}&=\mathbb{E}\left\lbrace\mathrm{tr}\left[\left(\mathbf{H}_{\mathrm{UL}}^T\mathbf{H}_{\mathrm{UL}}^*\right)^{-1}\right] \right\rbrace\overset{\equlabel}{=}\frac{M\mathrm{tr}\left\lbrace(\boldsymbol{\mathcal{B}}\mathbf{\Phi}^2\boldsymbol{\mathcal{B}}^{*})^{-1}\right\rbrace}{\mathrm{tr}\left\lbrace\mathbf{RR}^*\right\rbrace(M-K)},
	\end{split}
\end{equation}
where the step $(a)$ is because $(\mathbf{H}_{\mathrm{UL}}^T\mathbf{H}_{\mathrm{UL}}^*)^{-1}$ obeys the inverse-Wishart distribution denoted as $(\mathbf{H}_{\mathrm{UL}}^T\mathbf{H}_{\mathrm{UL}}^*)^{-1}\sim\mathcal{W}^{-1}(M(\boldsymbol{\mathcal{B}}\mathbf{\Phi}^2\boldsymbol{\mathcal{B}}^{*}\mathrm{tr}\left\lbrace\mathbf{RR}^*\right\rbrace)^{-1},K,M)$.

According to \cite[Eq. (14)]{Wei2016Mutual} and \cite[Eq. (40)]{Nie2020Nie2020Relaying}, we further approximate $ ({\mathbf{H}}_{\mathrm{UL}}^*{\mathbf{H}}_{\mathrm{UL}}^{T})^{-1} $ into a diagonal matrix as
\begin{equation}
	({\mathbf{H}}_{\mathrm{UL}}^T{\mathbf{H}}_{\mathrm{UL}}^{*})^{-1}\approx \frac{1}{\mathrm{tr}\left\lbrace\mathbf{RR}^*\right\rbrace}\mathrm{diag}\left(\frac{1}{|b_1|^2\phi_1^2},\cdots,\frac{1}{|b_K|^2\phi_K^2}\right).
	\label{eq:inversediag}
\end{equation}

Then, the transmitted signal $x_{\mathrm{b},m}$ of the $m$-th antenna with the ZF precoding can be rewritten as
\begin{equation}
	x_{\mathrm{b},m}=[\mathbf{W}]_{m\cdot}\mathbf{s}=\frac{r_m^*}{\mathrm{tr}\left\lbrace\mathbf{RR}^*\right\rbrace\sqrt{\beta_{\mathrm{ZF}}}}\sum_{i=1}^{K}\frac{h_{m,i}^*s_i}{b_i\phi_k}.
\end{equation}
The variance of $x_{\mathrm{b},m}$ can be derived as
\begin{equation}
	\begin{split}
		\sigma_{x,m}^2&=\mathbb{E}\left\lbrace|x_{\mathrm{b},m}|^2\right\rbrace\\
		&=\frac{\mathbb{E}\left\lbrace|\sum_{i=1}^{K}h_{m,i}^*s_ib_i^{-1}\phi_i^{-1}|^2\right\rbrace|r_m|^2}{(\mathrm{tr}\left\lbrace\mathbf{RR}^*\right\rbrace)^2\beta_{\mathrm{ZF}}}\\
		&=\frac{(M-K)\rho_t|r_m|^2\sum_{i=1}^{K}|h_{m,i}|^2|b_i\phi_i|^{-2}}{M\mathrm{tr}\left\lbrace\mathbf{RR}^*\right\rbrace\sum_{i=1}^{K}|b_i\phi_i|^{-2}}\\
		&\overset{\equlabel}{=}\frac{|r_m|^2\rho_{\mathrm{t}}}{\mathrm{tr}\left\lbrace\mathbf{RR}^*\right\rbrace},
	\end{split}
	\label{eq:sigmaxZF}
\end{equation}
where $(b)$ holds due to LLN. By substituting \eqref{eq:sigmaxZF} into $\mu(x)$, the closed-form expression of the linear scalar $ g_{\mathrm{ZF},m} $ of ZF can be denoted as
\begin{equation}
	\begin{split}
		g_{\mathrm{ZF},m}=t_m\mu\left(\frac{A_{\mathrm{Ast},m}\sqrt{\mathrm{tr}\left\lbrace\mathbf{RR}^*\right\rbrace}}{\sqrt{|r_m|^2\rho_{\mathrm{t}}}}\right).
	\end{split}
	\label{eq:gzfmdef}
\end{equation}
The closed-form expression of the variance $ \sigma_{\mathrm{d},m}^2 $ of the nonlinear distortion $ d_m $ of ZF can be denoted as
\begin{equation}
	\sigma_{\mathrm{ZF},m}^2=|t_m|^2\lambda_m\left(\sqrt{\frac{|r_m|^2\rho_{\mathrm{t}}}{\mathrm{tr}\left\lbrace\mathbf{RR}^*\right\rbrace}}\right).
	\label{eq:sigmaZFmdef}
\end{equation}

The effective downlink channel gain $ \mathbf{H}_{\mathrm{eq}}=\mathbf{H}_{\mathrm{DL}}\mathbf{W} $ for ZF can be denoted as
\begin{equation}
	\begin{split}
		\mathbf{H}_{\mathrm{eq}}&=\frac{1}{\sqrt{\beta_{\mathrm{ZF}}}}\mathbf{H}_{\mathrm{DL}}\mathbf{H}_{\mathrm{UL}}^*(\mathbf{H}_{\mathrm{UL}}^T\mathbf{H}_{\mathrm{UL}}^*)^{-1}\\
		&=\frac{1}{\sqrt{\beta_{\mathrm{ZF}}}}\mathbf{U}\bar{\mathbf{H}}\mathbf{R}^{-1}\mathbf{G}_{\mathrm{ZF}}\bar{\mathbf{H}}^H(\bar{\mathbf{H}}\bar{\mathbf{H}}^H)^{-1}\boldsymbol{\mathcal{B}}^{-1},
	\end{split}
\end{equation}
where $ \bar{\mathbf{H}}=\mathbf{HR}$, and $\mathbf{G}_{\mathrm{ZF}}=\mathrm{diag}(g_{\mathrm{ZF},1},\cdots,g_{\mathrm{ZF},M})$. To accurately derive the SINDR, we further denoted $\mathbf{R}^{-1}\mathbf{G}_{\mathrm{ZF}}=\alpha\mathbf{I}_{M}+\mathbf{\Delta}_{\mathrm{rg}}$, where $\alpha$ is the arithmetic mean of $g_m/r_m$, i.e., $\alpha=\frac{1}{M}\sum_{m=1}^{M}g_{\mathrm{ZF},m}/r_m$, $\mathbf{\Delta }=\mathrm{diag}(\Delta_1,\cdots,\Delta_M)$, the arithmetic mean of $\Delta_m$ is zero and the variance of $\Delta_m$ is equal to the variance of $g_{\mathrm{ZF},m}$. Then, the effective downlink channel can be further given by
\begin{equation}
	\begin{split}
		\mathbf{H}_{\mathrm{eq}}&=\frac{1}{\sqrt{\beta_{\mathrm{ZF}}}}\mathbf{U}\left[\bar{\mathbf{H}}\mathbf{\Delta}_{\mathrm{rg}}\bar{\mathbf{H}}^H(\bar{\mathbf{H}}\bar{\mathbf{H}}^H)^{-1}+\alpha\right]\boldsymbol{\mathcal{B}}^{-1}\\
		&=\frac{1}{\sqrt{\beta_{\mathrm{ZF}}}}\mathbf{U}\left[\frac{1}{\mathrm{tr}\left\lbrace\mathbf{RR}^*\right\rbrace}\bar{\mathbf{H}}\mathbf{\Delta}_{\mathrm{rg}}\bar{\mathbf{H}}^H \vphantom{\frac{1}{\mathrm{tr}\left\lbrace\mathbf{RR}^*\right\rbrace}\mathrm{diag}\left(\frac{1}{|b_1|^2\phi_1^2},\cdots,\frac{1}{|b_K|^2\phi_K^2}\right)}\mathrm{diag}\left(\frac{1}{\phi_1^2},\cdots,\frac{1}{\phi_K^2}\right)+\alpha\mathbf{I}_K\right]\boldsymbol{\mathcal{B}}^{-1},
	\end{split}
	\label{eq:zfchannelgain}
\end{equation}

In light of \eqref{eq:zfchannelgain}, the statistical effective channel gain $\mathbb{E}\left\lbrace h_{\mathrm{eq},k,k} \right\rbrace$ can be derived as
\begin{equation}
	\begin{split}
		\mathbb{E}\left\lbrace h_{\mathrm{eq},k,k}\right\rbrace&=\frac{u_k}{\sqrt{\beta_{\mathrm{ZF}}}b_k}\mathbb{E}\left\lbrace\frac{\bar{\mathbf{h}}_k\mathbf{\Delta}_{\mathrm{rg}} \bar{\mathbf{h}}_k^H}{\mathrm{tr}\left\lbrace \mathbf{RR}^*\right\rbrace\phi_k^2}+\alpha\right\rbrace\\
		&=\frac{u_k\mathrm{tr}\left\lbrace \mathbf{G}_{\mathrm{ZF}}\mathbf{R}^*-\alpha\mathbf{R}\mathbf{R}^* \right\rbrace}{\sqrt{\beta_{\mathrm{ZF}}}b_k\mathrm{tr}\left\lbrace\mathbf{RR}^*\right\rbrace}+\frac{\alpha u_k}{\sqrt{\beta_{\mathrm{ZF}}}b_k}\\
		&=\frac{u_k\mathrm{tr}\left\lbrace \mathbf{G}_{\mathrm{ZF}}\mathbf{R}^* \right\rbrace}{\sqrt{\beta_{\mathrm{ZF}}}b_k\mathrm{tr}\left\lbrace\mathbf{RR}^*\right\rbrace}.
	\end{split}
\end{equation}
Then, the power of the effective signal received at the $k$-th UE can be given by
\begin{equation}
	\begin{split}
		\varUpsilon_{\mathrm{ZF},k}^{\mathrm{ES}}&=a_0\rho_{\mathrm{t}}|\mathbb{E}\left\lbrace h_{\mathrm{eq},k,k}\right\rbrace|^2\\
		&=\frac{a_0\rho_{\mathrm{t}}(M-K)|u_k\mathrm{tr}\left\lbrace\mathbf{G}_{\mathrm{ZF}}\mathbf{R}^*\right\rbrace|^2}{M|b_k|^2\mathrm{tr}\left\lbrace\boldsymbol{(\mathcal{B}\Phi^2\mathcal{B}^*)^{-1}}\right\rbrace\mathrm{tr}\left\lbrace\mathbf{RR}^*\right\rbrace}.
	\end{split}
\end{equation}

The second moment of the statistical effective channel gain of ZF can be denoted as
\begin{equation}
	\begin{split}
		&\quad\ \mathbb{E}\left\lbrace |h_{\mathrm{eq},k,k}|^2 \right\rbrace\\
		&=\frac{|u_k|^2}{\beta_{\mathrm{ZF}}|b_k|^2}\mathbb{E}\left\lbrace\left|\frac{\bar{\mathbf{h}}_k\mathbf{\Delta}_{\mathrm{rg}} \bar{\mathbf{h}}_k^H}{\mathrm{tr}\left\lbrace \mathbf{RR}^*\right\rbrace\phi_k^2}+\alpha\right|^2\right\rbrace\\
		&=\frac{|u_k|^2}{\beta_{\mathrm{ZF}}|b_k|^2}\left[\frac{\mathbb{E}\left\lbrace|\bar{\mathbf{h}}_k\mathbf{\Delta}_{\mathrm{rg}} \bar{\mathbf{h}}_k^H|^2\right\rbrace}{(\mathrm{tr}\left\lbrace \mathbf{RR}^*\right\rbrace)^2\phi_k^4}+\frac{\mathbb{E}\left\lbrace \bar{\mathbf{h}}_k\mathbf{\Delta}_{\mathrm{rg}} \bar{\mathbf{h}}_k^H\right\rbrace}{\mathrm{tr}\left\lbrace \mathbf{RR}^*\right\rbrace\phi_k^2}+\frac{\mathbb{E}\left\lbrace \bar{\mathbf{h}}_k\mathbf{\Delta}_{\mathrm{rg}}^* \bar{\mathbf{h}}_k^H\right\rbrace}{\mathrm{tr}\left\lbrace \mathbf{RR}^*\right\rbrace\phi_k^2}+|\alpha|^2\right]\\
		&=\frac{|u_k|^2}{\beta_{\mathrm{ZF}}|b_k|^2}\left[\frac{\mathrm{tr}\left\lbrace(\mathbf{G}_{\mathrm{ZF}}-\alpha\mathbf{R})\mathbf{RR}^*(\mathbf{G}_{\mathrm{ZF}}^*-\alpha\mathbf{R}^*) \right\rbrace}{(\mathrm{tr}\left\lbrace \mathbf{RR}^*\right\rbrace)^2}+\frac{2\mathrm{Re}\left\lbrace\mathrm{tr} [(\mathbf{G}_{\mathrm{ZF}}^*-\alpha\mathbf{R}^*)\mathbf{R}]\right\rbrace}{\mathrm{tr}\left\lbrace \mathbf{RR}^*\right\rbrace}+|\alpha|^2\right.\\
		&\left.\quad+\frac{\mathrm{tr}\left\lbrace(\mathbf{G}_{\mathrm{ZF}}-\alpha\mathbf{R})\mathbf{R}^* \right\rbrace\mathrm{tr}\left\lbrace(\mathbf{G}_{\mathrm{ZF}}^*-\alpha\mathbf{R}^*)\mathbf{R} \right\rbrace}{(\mathrm{tr}\left\lbrace \mathbf{RR}^*\right\rbrace)^2}\right]\\
		&=\frac{|u_k|^2\mathrm{tr}\left\lbrace(\mathbf{G}_{\mathrm{ZF}}-\alpha\mathbf{R})(\mathbf{G}_{\mathrm{ZF}}^*-\alpha\mathbf{R}^*) \right\rbrace}{M\beta_{\mathrm{ZF}}|b_k|^2\mathrm{tr}\left\lbrace \mathbf{RR}^*\right\rbrace}+\frac{|u_k\mathrm{tr}\left\lbrace \mathbf{G}_{\mathrm{ZF}}\mathbf{R}^* \right\rbrace|^2}{M\beta_{\mathrm{ZF}}|b_k|^2(\mathrm{tr}\left\lbrace \mathbf{RR}^*\right\rbrace)^2}.
	\end{split}
\end{equation}

Hence, the power of the self-interference can be given by
\begin{equation}
	\begin{split}
		\varUpsilon_{\mathrm{ZF},k}^{\mathrm{SI}}&=a_0\rho_{\mathrm{t}}\mathrm{Var}\left\lbrace h_{\mathrm{eq},k,k}\right\rbrace\\
		&=a_0\rho_{\mathrm{t}}[\mathbb{E}\left\lbrace |h_{\mathrm{eq},k,k}|^2\right\rbrace-|\mathbb{E}\left\lbrace h_{\mathrm{eq},k,k}\right\rbrace|^2]\\
		&=\frac{a_0\rho_{\mathrm{t}}|u_k|^2\mathrm{tr}\left\lbrace(\mathbf{G}_{\mathrm{ZF}}-\alpha\mathbf{R})(\mathbf{G}_{\mathrm{ZF}}^*-\alpha\mathbf{R}^*) \right\rbrace}{(M-K)^{-1}M^2|b_k|^2\mathrm{tr}\left\lbrace(\boldsymbol{\mathcal{B}}\mathbf{\Phi}^2\boldsymbol{\mathcal{B}}^*)^{-1}\right\rbrace}.
	\end{split}
\end{equation}

Further, the power of the multi-user interference of ZF can be derived as
\begin{equation}
	\begin{split}
		\varUpsilon_{\mathrm{ZF},k}^{\mathrm{MUI}}&=a_0\rho_{\mathrm{t}}\sum_{i\neq k}^{K}\mathbb{E}\left\lbrace |h_{\mathrm{eq},k,i}|^2\right\rbrace\\
		&=\frac{a_0\rho_{\mathrm{t}}|u_k|^2}{\beta_{\mathrm{ZF}}}\sum_{i\neq k}^{K}\mathbb{E}\left\lbrace\left|\frac{\bar{\mathbf{h}}_k\mathbf{\Delta}_{\mathrm{rg}} \bar{\mathbf{h}}_i^H}{\mathrm{tr}\left\lbrace \mathbf{RR}^*\right\rbrace b_i\phi_i^2}\right|^2\right\rbrace\\
		&=\sum_{i\neq k}^{K}\frac{a_0\rho_{\mathrm{t}}|u_k|^2\phi_k^2\mathrm{tr}\left\lbrace (\mathbf{G}_{\mathrm{ZF}}-\alpha\mathbf{R})(\mathbf{G}_{\mathrm{ZF}}-\alpha\mathbf{R})^*\right\rbrace}{(M-K)^{-1}M^2|u_k^{-1}b_i|^2\phi_i^2\mathrm{tr}\left\lbrace(\boldsymbol{\mathcal{B}}\mathbf{\Phi}^2\boldsymbol{\mathcal{B}}^*)^{-1}\right\rbrace}.
	\end{split}
\end{equation}

Finally, the power of the nonlinear distortion can be evaluated as
\begin{equation}
	\varUpsilon_{\mathrm{ZF},k}^{\mathrm{NLI}}=a_0|u_k|^2\phi_k^2\mathbb{E}\left\lbrace\mathbf{h}_k\mathbf{\Sigma}_{\mathrm{ZF}}\mathbf{h}^H\right\rbrace=a_0|u_k|^2\phi_k^2\mathrm{tr}\left\lbrace\mathbf{\Sigma}_{\mathrm{ZF}}\right\rbrace,
\end{equation}
where $\mathbf{\Sigma}_{\mathrm{ZF}}=\mathrm{diag}(\sigma_{\mathrm{ZF},1}^2,\cdots,\sigma_{\mathrm{ZF},M}^2)$.

\section{Proof of Proposition \ref{coro:sinrlimitinterzf}}\label{proof:sinrlimitinterzf}
As the HPA works with the large IBO state, the average power of the input signal is less than the saturation level, i.e., $\frac{\sigma_{\mathrm{x},m}^2}{A_{\mathrm{sat},m}^2}<1 $. Hence, we use Taylor series centered at $\epsilon$ to approximate the function $f(x)=\mu(1/x)$, where $\epsilon$ is a very small positive constant. By utilizing the Taylor expansion,  the function $f(x)$ denoted as
\begin{equation}
	\begin{split}
		f_{\epsilon}\left(x\right)&=f(\epsilon)-f'(\epsilon)(x-\epsilon)+\frac{f''(\epsilon)}{2}(x-\epsilon)^2+\frac{f'''(\epsilon)}{6}(x-\epsilon)^3+\mathcal{O}((x-\epsilon)^3).
	\end{split}
	\label{eq:asymptmu1}
\end{equation}
When $\epsilon$ approaches to $0$, $f(x)$ can be given by
\begin{equation}
	f_0(x)=\lim_{\epsilon\rightarrow 0}f(x)\approx 1-x^2+\mathcal{O}(x^3).
\end{equation}
By substituting $\sigma_{\mathrm{x},m}^2 $ denoted in \eqref{eq:sigmaxZF} into $f_0(x)$, $g_{\mathrm{ZF},m}$ can be approximated as
\begin{equation}
	\begin{split}
		g_{\mathrm{ZF},m}&=t_m\mu\left(\frac{A_{\mathrm{Ast},m}\sqrt{\mathrm{tr}\left\lbrace\mathbf{RR}^*\right\rbrace}}{\sqrt{|r_m|^2\rho_{\mathrm{t}}}}\right)\approx t_m\left(1-\frac{|r_m|^2\rho_tA_{\mathrm{sat},m}^{-2}}{\mathrm{tr}\left\lbrace\mathbf{RR}^*\right\rbrace}\right).
	\end{split}
	\label{eq:gzflimitinter}
\end{equation}

Similarly, by exploiting the Taylor expansion of $\lambda_m(1/x)$, the variance of the nonlinear distortion can be approximated as
\begin{equation}
	\sigma_{\mathrm{ZF},m}^2\approx \frac{|r_m|^6\rho_t^3A_{\mathrm{sat},m}^{-6}}{2(\mathrm{tr}\left\lbrace \mathbf{RR}^*\right\rbrace)^3}.
	\label{eq:sigmazflimitinter}
\end{equation}

By substituting \eqref{eq:gzflimitinter} into \eqref{eq:poweres}, the power of effective signal can be further given by
\begin{equation}
	\begin{split}
		\bar{\varUpsilon}_{\mathrm{ZF},k}^{\mathrm{ES}}&=a_0\rho_{\mathrm{t}}\frac{\left|\sum_{m=1}^{M}t_mr_m^*\left(1-\frac{|r_m|^2\rho_tA_{\mathrm{sat},m}^{-2}}{\mathrm{tr}\left\lbrace\mathbf{RR}^*\right\rbrace}\right)\right|^2}{(M-K)^{-1}M\mathrm{tr}\left\lbrace \mathbf{\Phi}^{-2}\right\rbrace\sum_{m=1}^{M}|r_m|^2}\\
		&\overset{\equlabel}{=}a_0\rho_{\mathrm{t}}\frac{(M-K)|\mathbb{E}\left\lbrace t_mr_m^*\right\rbrace|^2}{\mathrm{tr}\left\lbrace \mathbf{\Phi}^{-2}\right\rbrace\mathbb{E}\left\lbrace |r_m|^2\right\rbrace}\left(1-\frac{\mathbb{E}\left\lbrace |a_m|^{-2}\right\rbrace}{M\rho_{\mathrm{t}}^{-1}A_{\mathrm{sat}}^2}\right)^2\\
		&\overset{\equlabel}{\approx} a_0\rho_{\mathrm{t}}\frac{(M-K)|\mathbb{E}\left\lbrace t_mr_m^*\right\rbrace|^2}{\mathrm{tr}\left\lbrace \mathbf{\Phi}^{-2}\right\rbrace\mathbb{E}\left\lbrace |r_m|^2\right\rbrace}\left(1-\frac{2\mathbb{E}\left\lbrace |a_m|^{-2}\right\rbrace}{M\rho_{\mathrm{t}}^{-1}A_{\mathrm{sat}}^2}\right),
		\label{eq:epoweres}
	\end{split}
\end{equation}
where $(a)$ holds due to LLN, and $(c)$ is conditioned on ignoring the high-order infinitesimality of $\rho_{\mathrm{t}}/(MA_{\mathrm{sat}})$. Substituting \eqref{eq:gzflimitinter} into \eqref{eq:powersi} and \eqref{eq:powermui}, the sum of self-interference and multi-user interference can be further denoted as
\begin{equation}
	\begin{split}
		\bar{\varUpsilon}_{\mathrm{ZF},k}^{\mathrm{SI}}&=a_0\rho_{\mathrm{t}}\frac{M-K}{M^2\mathrm{tr}\left\lbrace \mathbf{\Phi}^{-2}\right\rbrace}\sum_{m=1}^{M}|g_{\mathrm{ZF},m}-\alpha r_m|^2\\
		&=a_0\rho_{\mathrm{t}}\frac{M-K}{M^2\mathrm{tr}\left\lbrace \mathbf{\Phi}^{-2}\right\rbrace}\sum_{m=1}^{M}\left[|g_{\mathrm{ZF},m}|^2+|\alpha r_m|^2-2\mathrm{Re}\left\lbrace g_{\mathrm{ZF},m}\alpha^*r_m^* \right\rbrace\right]\\
		&\overset{\equlabel}{=}\frac{a_0\rho_{\mathrm{t}}(M-K)}{M\mathrm{tr}\left\lbrace \mathbf{\Phi}^{-2}\right\rbrace}\left[\mathbb{E}\left\lbrace |t_m|^2\right\rbrace+\left|\mathbb{E}\left\lbrace \frac{t_m}{r_m}\right\rbrace\right|^2\mathbb{E}\left\lbrace |r_m|^2\right\rbrace\right.\\
		&\quad\left.-2\mathrm{Re}\left(\mathbb{E}\left\lbrace \frac{t_m}{r_m}\right\rbrace \mathbb{E}\left\lbrace t_m^*r_m\right\rbrace\right)\right]\left(1-\frac{2\rho_{\mathrm{t}}\mathbb{E}\left\lbrace a_m^{-2}\right\rbrace}{MA_{\mathrm{sat}}^2}\right),
	\end{split}
	\label{eq:epowersi}
\end{equation}
where $(c)$ holds due to LLN and ignoring the high-order infinitesimality of $\rho_{\mathrm{t}}/(MA_{\mathrm{sat}})$. Similarly, the power of multi-user interference can be rewritten as
\begin{equation}
	\begin{split}
		\varUpsilon_{\mathrm{ZF},k}^{\mathrm{MUI}}&=\frac{a_0\rho_{\mathrm{t}}(M-K)\phi_k^2\sum_{i\neq k}^{K}\phi_i^{-2}}{M\mathrm{tr}\left\lbrace \mathbf{\Phi}^{-2}\right\rbrace}\left[\mathbb{E}\left\lbrace |t_m|^2\right\rbrace2\mathrm{Re}\left(\mathbb{E}\left\lbrace \frac{t_m}{r_m}\right\rbrace \mathbb{E}\left\lbrace t_m^*r_m\right\rbrace\right)\right.\\
		&\quad\left.  +\left|\mathbb{E}\left\lbrace \frac{t_m}{r_m}\right\rbrace\right|^2\mathbb{E}\left\lbrace |r_m|^2\right\rbrace \right]\left(1-\frac{2\rho_{\mathrm{t}}\mathbb{E}\left\lbrace a_m^{-2}\right\rbrace}{MA_{\mathrm{sat}}^2}\right).
	\end{split}
	\label{eq:epowermui}
\end{equation}

According to \eqref{eq:sigmazflimitinter}, $ \sigma_{\mathrm{d},m}^2 $ is a higher-order infinitesimality of $\frac{\rho_t}{MA_{\mathrm{sat}}^2}$ when $\frac{\rho_t}{MA_{\mathrm{sat}}^2}<1$, the impact of $ \sigma_{\mathrm{d},m}^2 $ is slight and can be ignored. Then, by substituting \eqref{eq:epoweres}, \eqref{eq:epowersi}, and \eqref{eq:epowermui} into \eqref{eq:sinrzf} and exploiting the statistic properties of $t_m$, $r_m$, and $a_m$, SINR with the large IBO can be given by \eqref{eq:highpowerZF}. Therefore, Proposition \ref{coro:sinrlimitinterzf} holds.

\bibliographystyle{IEEEtran}
\bibliography{reference}

\begin{thebibliography}{10}
\providecommand{\url}[1]{#1}
\csname url@samestyle\endcsname
\providecommand{\newblock}{\relax}
\providecommand{\bibinfo}[2]{#2}
\providecommand{\BIBentrySTDinterwordspacing}{\spaceskip=0pt\relax}
\providecommand{\BIBentryALTinterwordstretchfactor}{4}
\providecommand{\BIBentryALTinterwordspacing}{\spaceskip=\fontdimen2\font plus
\BIBentryALTinterwordstretchfactor\fontdimen3\font minus
  \fontdimen4\font\relax}
\providecommand{\BIBforeignlanguage}[2]{{%
\expandafter\ifx\csname l@#1\endcsname\relax
\typeout{** WARNING: IEEEtran.bst: No hyphenation pattern has been}%
\typeout{** loaded for the language `#1'. Using the pattern for}%
\typeout{** the default language instead.}%
\else
\language=\csname l@#1\endcsname
\fi
#2}}
\providecommand{\BIBdecl}{\relax}
\BIBdecl

\bibitem{Larsson2014Massive}
E.~G. Larsson, O.~Edfors, F.~Tufvesson, and T.~L. Marzetta, ``{Massive MIMO for
  next generation wireless systems},'' \emph{IEEE Commun. Mag.}, vol.~52,
  no.~2, pp. 186--195, Feb. 2014.

\bibitem{Larsson2015Fundamentals}
T.~L. Marzetta, E.~G. Larsson, H.~Yang, and H.~Q. Ngo, \emph{Fundamentals of
  Massive MIMO}.\hskip 1em plus 0.5em minus 0.4em\relax Cambridge University
  Press, Nov. 2016.

\bibitem{Akyildiz20165G}
I.~F. Akyildiz, S.~Nie, S.~C. Lin, and M.~Chandrasekaran, ``{5G roadmap: 10 key
  enabling technologies},'' \emph{Comput. Networks}, vol. 106, pp. 17--48, Sep.
  2016.

\bibitem{Rusek2013Scaling}
F.~Rusek, D.~Persson, B.~K. Lau, E.~G. Larsson, T.~L. Marzetta, O.~Edfors, and
  F.~Tufvesson, ``{Scaling up MIMO : Opportunities and challenges with very
  large arrays},'' \emph{IEEE Signal Process. Mag.}, vol.~30, no.~1, pp.
  40--60, Jan. 2013.

\bibitem{Shan2018a}
C.~Shan, L.~Chen, X.~Chen, and W.~Wang, ``{A general matched filter design for
  reciprocity calibration in multiuser massive MIMO systems},'' \emph{IEEE
  Trans. Veh. Technol.}, vol.~67, no.~9, pp. 8939--8943, Sep. 2018.

\bibitem{Shan2017Performance}
C.~Shan, Y.~Zhang, L.~Chen, X.~Chen, and W.~Wang, ``{Performance Analysis of
  Large Scale Antenna System with Carrier Frequency Offset, Quasi-Static
  Mismatch and Channel Estimation Error},'' \emph{IEEE Access}, vol.~5, pp.
  26\,135--26\,145, Nov. 2017.

\bibitem{Mi2017Massive}
D.~Mi, M.~Dianati, L.~Zhang, S.~Muhaidat, and R.~Tafazolli, ``{Massive MIMO
  Performance With Imperfect Channel Reciprocity and Channel Estimation
  Error},'' \emph{IEEE Trans. Commun.}, vol.~65, no.~9, pp. 3734--3749, Sep.
  2017.

\bibitem{Zhang2015Large}
W.~Zhang, H.~Ren, C.~Pan, M.~Chen, R.~C. {De Lamare}, B.~Du, and J.~Dai,
  ``{Large-scale antenna systems with UL/DL hardware mismatch: Achievable rates
  analysis and calibration},'' \emph{IEEE Trans. Commun.}, vol.~63, no.~4, pp.
  1216--1229, Apr. 2015.

\bibitem{Raeesi2018Performance}
O.~Raeesi, A.~Gokceoglu, Y.~Zou, E.~Bj{\"{o}}rnson, and M.~Valkama,
  ``{Performance Analysis of Multi-User Massive MIMO Downlink under Channel
  Non-Reciprocity and Imperfect CSI},'' \emph{IEEE Trans. Commun.}, vol.~66,
  no.~6, pp. 2456--2471, Jun. 2018.

\bibitem{Wei2016Impact}
H.~Wei, D.~Wang, J.~Wang, and X.~You, ``{Impact of RF mismatches on the
  performance of massive MIMO systems with ZF precoding},'' \emph{Sci. China
  Inf. Sci.}, vol.~59, no.~2, pp. 1--14, Jan. 2016.

\bibitem{Jiang2015MIMO}
X.~Jiang, M.~Cirkic, F.~Kaltenberger, E.~G. Larsson, L.~Deneire, and R.~Knopp,
  ``{MIMO-TDD reciprocity under hardware imbalances: Experimental results},''
  \emph{IEEE Int. Conf. Commun.}, vol. 2015-Septe, pp. 4949--4953, Jun. 2015.

\bibitem{Nie2019A}
R.~Nie, L.~Chen, C.~Shan, and X.~Chen, ``{A decentralized reciprocity
  calibration approach for cooperative MIMO},'' \emph{IEEE Access}, vol.~7, pp.
  1560--1569, Dec. 2019.

\bibitem{Calibration2001Automatic}
K.~Nishimori, K.~Cho, Y.~Takatori, and T.~Hori, ``{Automatic calibration method
  using transmitting signals of an adaptive array for TDD systems},''
  \emph{IEEE Trans. Veh. Technol.}, vol.~50, no.~6, pp. 1636--1640, Nov. 2001.

\bibitem{Bourdoux2003Non}
A.~Bourdoux, B.~Come, and N.~Khaled, ``{Non-reciprocal transceivers in
  OFDM/SDMA systems: Impact and mitigation},'' in \emph{Proc. IEEE Radio Wirel.
  Conf. (RAWCON)}, Aug. 2003, pp. 183--186.

\bibitem{Liu2004OFDM}
J.~Liu, A.~Bourdoux, J.~Craninckx, P.~Wambacq, B.~C{\^{o}}me, S.~Donnay, and
  A.~Barel, ``{OFDM-MIMO WLAN AP front-end gain and phase mismatch
  calibration},'' in \emph{Proc. 2004 IEEE Radio Wirel. Conf. (RAWCON)}, Feb.
  2004, pp. 151--154.

\bibitem{JianLiu2006A}
{Jian Liu}, G.~Vandersteen, J.~Craninckx, M.~Libois, M.~Wouters, F.~Petre, and
  A.~Barel, ``{A novel and low-cost analog front-end mismatch calibration
  scheme for MIMO-OFDM WLANs},'' in \emph{Proc. 2006 IEEE Radio Wireless
  Symp.}, Apr. 2006, pp. 219--222.

\bibitem{Benzin2017Internal}
A.~Benzin and G.~Caire, ``{Internal Self-Calibration Methods for Large Scale
  Array Transceiver Software-Defined Radios},'' in \emph{Proc. Int. ITG
  Workshop Smart Antennas (WSA)}, Mar. 2017, pp. 49--56.

\bibitem{Luo2019Massive}
X.~Luo, F.~Yang, and H.~Zhu, ``{Massive MIMO Self-Calibration: Optimal
  Interconnection for Full Calibration},'' \emph{IEEE Trans. Veh. Technol.},
  vol.~68, no.~11, pp. 10\,357--10\,371, Nov. 2019.

\bibitem{Guillaud2005A}
M.~Guillaud, D.~Slock, and R.~Knopp, ``{A practical method for wireless channel
  reciprocity exploitation through relative calibration},'' in \emph{Proc. 8th
  Int. Symp. Signal Process. Applic. (ISSPA)}, {Intergovernmental Panel on
  Climate Change}, Ed., vol.~1, no.~9, Jan. 2005, pp. 403--406.

\bibitem{Kaltenberger2010Relative}
F.~Kaltenberger, J.~Haiyong, M.~Guillaud, and R.~Knopp, ``{Relative channel
  reciprocity calibration in MIMO/TDD systems},'' in \emph{Proc. Futur. Netw.
  Mob. Summit}, Mar. 2011, pp. 1--10.

\bibitem{Kouassi2012Estimation}
B.~Kouassi, I.~Ghauri, and L.~Deneire, ``{Estimation of Time-Domain Calibration
  Parameters to Restore MIMO-TDD Channel Reciprocity},'' in \emph{Proc. 7th
  Int. ICST Conf. Cogn. Radio Oriented Wirel. Networks Commun.}, Jul. 2012, pp.
  254--258.

\bibitem{r1092359hardware}
R1-092359, ``{Hardware calibration requirement for dual layer beamforming},''
  Huawei, Los Angeles, CA, USA, 3GPP RAN1 57, Jun. 2009.

\bibitem{Shepard2012Argos}
C.~Shepard, H.~Yu, N.~Anand, E.~Li, T.~Marzetta, R.~Yang, and L.~Zhong,
  ``{Argos: practical many-antenna base stations},'' in \emph{Proc. 18th Annu.
  Int. Conf. Mobile Comput. Networking (Mobicom)}, Aug. 2012, pp. 53--64.

\bibitem{Wei2016Mutual}
H.~Wei, D.~Wang, H.~Zhu, J.~Wang, S.~Sun, and X.~You, ``{Mutual Coupling
  Calibration for Multiuser Massive MIMO Systems},'' \emph{IEEE Trans. Wireless
  Commun.}, vol.~15, no.~1, pp. 606--619, Jan. 2016.

\bibitem{Rogalin2013Hardware}
R.~Rogalin, O.~Y. Bursalioglu, H.~C. Papadopoulos, G.~Caire, and A.~F. Molisch,
  ``{Hardware-impairment compensation for enabling distributed large-scale
  MIMO},'' in \emph{Proc. 2013 Inf. Theory Appl. Work. (ITA)}, Feb. 2013, pp.
  304--313.

\bibitem{Jiang2018A}
X.~Jiang, A.~Decurninge, K.~Gopala, F.~Kaltenberger, M.~Guillaud, S.~Member,
  D.~Slock, and L.~Deneire, ``{A Framework for Over-the-air Reciprocity
  Calibration for TDD Massive MIMO Systems},'' \emph{IEEE Trans. Wireless
  Commun.}, vol.~17, no.~9, pp. 5975--5990, Sep. 2018.

\bibitem{Nie2020Nie2020Relaying}
R.~Nie, L.~Chen, N.~Zhao, Y.~Chen, F.~R. Yu, and G.~Wei, ``{Relaying Systems
  With Reciprocity Mismatch: Impact Analysis and Calibration},'' \emph{IEEE
  Trans. Commun.}, vol.~68, no.~7, pp. 4035--4049, Jul. 2020.

\bibitem{Guerreiro2018Analytical}
J.~Guerreiro, R.~Dinis, and P.~Montezuma, ``{Analytical Performance Evaluation
  of Precoding Techniques for Nonlinear Massive MIMO Systems with Channel
  Estimation Errors},'' \emph{IEEE Trans. Commun.}, vol.~66, no.~4, pp.
  1440--1451, Apr. 2018.

\bibitem{Balti2017Impact}
E.~Balti and M.~Guizani, ``{Impact of Non-Linear High-Power Amplifiers on
  Cooperative Relaying Systems},'' \emph{IEEE Trans. Commun.}, vol.~65, no.~10,
  pp. 4163--4175, Oct. 2017.

\bibitem{Yu2019Full}
C.~Yu, J.~Jing, H.~Shao, Z.~H. Jiang, P.~Yan, X.~W. Zhu, W.~Hong, and A.~Zhu,
  ``{Full-Angle Digital Predistortion of 5G Millimeter-Wave Massive MIMO
  Transmitters},'' \emph{IEEE Trans. Microw. Theory Techn.}, vol.~67, no.~7,
  pp. 2847--2860, Jul. 2019.

\bibitem{Liu2019Linearization}
X.~Liu, W.~Chen, L.~Chen, F.~M. Ghannouchi, and Z.~Feng, ``{Linearization for
  Hybrid Beamforming Array Utilizing Embedded Over-the-Air Diversity
  Feedbacks},'' \emph{IEEE Trans. Microw. Theory Techn.}, vol.~67, no.~12, pp.
  5235--5248, Dec. 2019.

\bibitem{Rowe1982Memoryless}
H.~E. Rowe, ``{Memoryless Nonlinearities With Gaussian Inputs: Elementary
  Results},'' \emph{Bell Syst. Tech. J.}, vol.~61, no.~7, pp. 1519--1526, Sep.
  1982.

\bibitem{KrishnanLinear2016}
R.~Krishnan, M.~R. Khanzadi, N.~Krishnan, Y.~Wu, A.~GraellAmat, T.~Eriksson,
  and R.~Schober, ``{Linear Massive MIMO Precoders in the Presence of Phase
  Noise - A Large-Scale Analysis},'' \emph{IEEE Trans. Veh. Technol.}, vol.~65,
  no.~5, pp. 3057--3071, May 2016.

\bibitem{Raich2004Orthogonal}
R.~Raich, H.~Qian, and G.~T. Zhou, ``{Orthogonal polynomials for power
  amplifier modeling and predistorter design},'' \emph{IEEE Trans. Veh.
  Technol.}, vol.~53, no.~5, pp. 1468--1479, Sep. 2004.

\bibitem{boyd2004convex}
S.~Boyd, S.~P. Boyd, and L.~Vandenberghe, \emph{Convex optimization}.\hskip 1em
  plus 0.5em minus 0.4em\relax Cambridge university press, 2004.

\bibitem{Powell1978A}
M.~Powell, ``{A fast algorithm for nonlinearly constrained optimization
  calculations},'' in \emph{Numerical analysis}.\hskip 1em plus 0.5em minus
  0.4em\relax Berlin, Germany: Springer-Verlag, 1978, vol. 630, pp. 144--157.

\end{thebibliography}

\end{document}